\documentclass[fleqn,usenatbib]{mnras}

\usepackage[T1]{fontenc}
\usepackage{ae,aecompl}

\usepackage{graphicx}	
\newsavebox\CBox

\usepackage{amsmath}	
\usepackage{amssymb}	

\usepackage{amsfonts}
\usepackage{bm}
\usepackage{hyperref}
\usepackage{threeparttable}
\usepackage{array}
\usepackage{multirow}
\usepackage{framed}
\usepackage{rotating}
\usepackage{xcolor}

\usepackage{fixltx2e}

\usepackage{newtxtext,newtxmath}

\newcommand{\pz}{\phantom{0}}
\newcommand{\pp}{\phantom{$+$}}

\title[Candidate PHz protoclusters and lensed galaxies]
{Candidate high-redshift protoclusters and lensed galaxies in the \textit{Planck\/} list of high-\textit{z} sources overlapping with \textit{Herschel}-SPIRE imaging}

\author[Lammers et al.]
{Caleb Lammers,$^{1, 2}$
Ryley Hill,$^{1}$
Seunghwan Lim,$^{1, 3}$
Douglas Scott,$^{1}$
Raoul Ca\~nameras,$^{4}$
Herv\'e Dole$^{5}$
\\
$^{1}$Department of Physics and Astronomy, University of British Columbia, 6225 Agricultural Road, Vancouver, V6T 1Z1, Canada\\
$^{2}$Department of Physics, University of Toronto, 60 St.\ George Street, Toronto, ON M5S 1A7, Canada\\
$^{3}$Canadian Institute for Theoretical Astrophysics, University of Toronto, 60 St.\ George Street, Toronto, ON M5S 3H8, Canada\\
$^{4}$Max-Planck-Institut f\"ur Astrophysik, Karl-Schwarzschild Str. 1, 85748 Garching, Germany\\
$^{5}$Universit\'e Paris-Saclay, CNRS, Institut d'astrophysique spatiale, 91405, Orsay, France
}

\pubyear{2022}

\begin{document}
\maketitle

\begin{abstract}
\noindent The {\it Planck\/} list of high-redshift source candidates (the PHz catalogue) contains 2151 peaks in the cosmic infrared background, unresolved by {\it Planck\/}'s 5\,arcmin beam. Follow-up spectroscopic observations have revealed that some of these objects are $z\,{\approx}\,2$ protoclusters and strong gravitational lenses, but an unbiased survey has not yet been carried out. To this end, we have used archival {\it Herschel}-SPIRE observations to study a uniformly-selected sample of 187 PHz sources. In contrast with follow-up studies that were biased towards bright, compact sources, we find that only one of our PHz sources is a bright gravitationally-lensed galaxy (peak flux ${\gtrsim}\,300$\,mJy), indicating that such objects are rarer in the PHz catalogue than previously believed ({<}\,1\,per cent). The majority of our PHz sources consist of many red, star-forming galaxies, demonstrating that typical PHz sources are candidate protoclusters. However, our new PHz sources are significantly less bright than found in previous studies and differ in colour, suggesting possible differences in redshift and star-formation rate. Nonetheless, 40 of our PHz sources contain ${>}\,3\,\sigma$ galaxy overdensities, comparable to the fraction of ${>}\,3\,\sigma$ overdensities found in earlier biased studies. We additionally use a machine-learning approach to identify less extreme (peak flux ${\sim}\,100$\,mJy) gravitationally-lensed galaxies among {\it Herschel}-SPIRE observations of PHz sources, finding a total of seven candidates in our unbiased sample, and 13 amongst previous biased samples. Our new uniformly-selected catalogues of ${>}\,3\,\sigma$ candidate protoclusters and strong gravitational lenses provide interesting targets for follow up with higher-resolution facilities, such as ALMA and \textit{JWST}.
\end{abstract}


\section{Introduction}
\label{introduction}
Wide sky surveys at millimetre and submillimetre (submm) wavelengths are now routinely carried out to map the cosmic microwave background (CMB) at arcminute resolution for cosmological studies. In the process, these surveys inevitably find the brightest foreground (i.e., between the Earth and the surface of last scattering at $z\,{\approx}\,1100$) millimetre and submm sources, which are themselves interesting for studying star-formation at various epochs. Some of these sources are bright star-forming regions in our Galaxy, while others are nearby galaxies or active galactic nuclei that can be filtered out with various selection criteria \citep[e.g.,][]{Planck2011_XVI,Planck2011_XXIII,planck2014-a37,PlanckPCCS22016,everett2020}; however, after follow-up observations with higher-resolution imaging, the remaining sources are now known to be strong gravitational lenses \citep[e.g.,][]{hezaveh2013,CanamerasPlanckGEMS2015,spilker2016} or overdensities of star-forming galaxies \citep[e.g.,][]{PlanckHerschel2015,Flores-CachoG952016,MillerSPT2349-562018,KneisslG732019,HillSPT2349-562020,wang2020,KoyamaG2372021, PollettaG2372021}.

Strong gravitationally-lensed galaxies are useful targets for studying galaxy evolution. The magnification provided by lensing allows for the study of high-$z$ galaxies in much more detail than would be otherwise possible. This has significantly contributed to our understanding of the physical mechanisms driving star-formation at high redshift, through detailed studies of faint emission lines and resolved star-forming regions down to sub-kpc scales \citep[e.g.,][]{EbbelsLensed1996, LivermoreLensed2012, MassardiLensed2018}. Lensing can also be used to probe the substructure of dark matter halos, a key technique for constraining the properties of dark matter \citep[e.g.,][]{MaoSchneiderLensed1998, DalalCDMSubstructure2002, VegettiCDMSubstructure2014}. To place tight constraints on dark matter properties, a large sample of bright gravitational lenses is required. However, currently only a few tens of sufficiently bright gravitationally-lensed systems are known. Previously, searching for such systems primarily relied on large optical surveys \citep[e.g.,][]{jacobs2019}, but recently attention has been turning to the submm due to the efficiency at identifying high-$z$ sources, the ease of distinguishing between the lensed galaxy and foreground lens, and the potential for follow up with the Atacama Large Millimeter/Submillimeter Array \citep[ALMA,][]{Blainsubmm2002, HezavehCDMSubstructure2016, RitondaleLensed2019}. Submm all-sky surveys, such as the three highest frequency-channel maps from the {\it Planck\/} High-frequency Instrument (HFI), provide the opportunity to greatly increase the known sample of gravitationally-lensed systems, as well as to study the brightest ones in detail \citep{CanamerasPlanckGEMS2015}.

Despite the advances made in understanding galaxy evolution, the formation of galaxy clusters is not yet well understood. Studies of the progenitors of galaxy clusters, termed `protoclusters,' can place important constraints on galaxy formation models and the ingredients going into cosmological simulations \citep{HarrisonColes2011, MillerSPT2349-562018}. Furthermore, protoclusters are believed to represent the phase at which galaxies in galaxy clusters formed most of their stars, contributing ${\gtrsim}\,20$\,per cent to the cosmic star-formation rate (SFR) at $z\,{>}\,2$ \citep{MadauDickinsonCosmicSF2014, GranatoSims2015, ChiangProtoclusters2017}. However, current galaxy formation models struggle to reproduce the SFRs observed in the most dense protoclusters \citep{HillSPT2349-562020, LimProtoclusters2020}. As such, the rare and extreme population of high-$z$ protoclusters can help to constrain models of galaxy evolution. Though difficult to identify, protoclusters are visible in the far-IR (FIR) and submm during this starbursting phase due to the copious amounts of dust-enshrouded star formation they host. Millimetre-wavelength data can select protoclusters at higher redshifts, but the {\it Planck\/} all-sky maps, which we focus on here, are effective for selecting the far-IR spectral maximum from $z\,{\approx}\,2$, when the global star-formation activity peaked \citep{PlanckHerschel2015, MacKenzieSCUBA2017}.

{\it Planck}'s ability to identify foreground objects of astrophysical significance was recognized early on. In particular, the Planck Collaboration released three catalogues of compact sources: the Planck Early Release Compact Source Catalogue (ERCSC; \citealt{PlanckERCSC2011}); the Planck Catalogue of Compact Sources (PCCS; \citealt{PlanckPCCS2014}); and the Second Planck Catalogue of Compact sources (PCCS2; \citealt{PlanckPCCS22016}). Several groups followed up these catalogues in search of high-$z$ clusters and protoclusters \citep{HerranzHerschelPlanck2013, ClementsHerschelPlanck2014, BaesHerschelPlanck2014, GreensladeHerschelPCCS2018}. Specifically, \citet{HerranzHerschelPlanck2013}, \citet{ClementsHerschelPlanck2014} and \citet{BaesHerschelPlanck2014} looked for overdensities of dusty star-forming galaxies in the ERCSC, using early {\it Herschel\/} wide-field surveys (H-ATLAS, HerMES and HeViCS, respectively). \citet{HerranzHerschelPlanck2013} discovered a $z\,{=}\,3.26$ candidate protocluster, while \citet{ClementsHerschelPlanck2014} found another four candidates. On the other hand, \citet{BaesHerschelPlanck2014} found no candidate protoclusters in the PCCS that fell in the HeViCS field. Most recently, \citet{GreensladeHerschelPCCS2018} found 27 candidate protoclusters in the PCCS and PCCS2 that fell within the 800\,deg$^{2}$ of available {\it Herschel\/} survey area.

In addition to the various compact source catalogues, the Planck Collaboration released a catalogue of 2151 potential high-$z$ sources (the Planck list of high-$z$ source candidates, or PHz) by looking at peaks in the cosmic infrared background \citep{PlanckPHz2016}. \citet{PlanckHerschel2015}, hereafter PlanckXXVII, performed a dedicated {\it Herschel}-SPIRE follow up of 228 {\it Planck}-selected high-$z$ candidates, consisting of red sources from the PCCS and sources selected in a similar manner to the PHz catalogue. PlanckXXVII found that 212 of their {\it Planck\/} sources are consistent with being overdensisities of high-$z$ star-forming galaxies (i.e., potential protoclusters) and 12 are strong gravitationally-lensed candidates that have all since been confirmed (one was previously discovered); the remaining four sources were dominated by Galactic cirrus. Many of the 212 PlanckXXVII sources are now confirmed submm overdensities based on follow-up studies \citep{MacKenzieSCUBA2017, FryeG1652019, MartinacheSpitzer2018, KneisslG732019, ChengSCUBA2020}, and the newly-discovered lensed galaxies have been the focus of a number of analyses \citep[e.g.,][]{CanamerasPlanckGEMS2015, NesvadbaGEMS2017, CanamerasGEMS2017, Canameras2GEMS2017, CanamerasGEMS2018}. Follow-up observations of individual PHz overdensities have now confirmed a handful of protoclusters \citep{Flores-CachoG952016,KneisslG732019, KoyamaG2372021, PollettaG2372021}. Despite the success of the PHz in identifying high-$z$ protoclusters and gravitationally-lensed galaxies, no additional comprehensive follow-up studies of PHz sources have been done. Furthermore, PlanckXXVII followed up a biased sample of PHz sources, focusing on compact, high signal-to-noise objects and a comprehensive, unbiased study has not yet been carried out.

While PlanckXXVII was able to establish the existence of strong gravitational lenses and protoclusters in the PHz catalogue, the fact that the follow-up observations were biased towards bright and compact sources meant that the underlying statistics of the catalogue could not be robustly determined. In order to correctly deduce the nature of the PHz catalogue, in this paper we investigate a sample of 187 unbiased {\it Herschel}-SPIRE observations of PHz sources that serendipitously fall within the 1270\,deg$^{2}$ coverage of the {\it Herschel} wide-field surveys, now combined into a single data release by the {\it Herschel} Extragalactic Legacy Project \citep[HELP;][]{shirley2021}. Additionally, we use our unbiased sample to produce sub-samples of the most statistically significant strong gravitational lenses and protocluster candidates that can be followed up with future high-resolution submm surveys, ultimately improving the sample sizes of these objects and enabling more detailed studies of dark matter halos, star formation and high-redshift clustering.

Section~\ref{datasets} describes the data used in this paper, including the PHz catalogue and maps obtained from HELP. In Section~\ref{observations} we describe our {\it Herschel}-SPIRE photometry methods and in Section~\ref{sourcestats} we present the statistics of the PHz HELP sources and compare them to results from PlanckXXVII. In Section~\ref{PC_GLcandidates}, we identify statistically-significant candidate protoclusters and gravitationally-lensed galaxies, and the discussion and conclusions are presented in Sections~\ref{discussion} and \ref{conclusion}, respectively. Throughout this paper, we assume a standard $\Lambda$CDM cosmology with parameters from \citet{PlanckParams2018}.

\section{Data}
\label{datasets}
Here we provide a brief overview of the PHz catalogue and the HELP data release. For more details on the PHz catalogue and HELP, see \citet{PlanckPHz2016} and \citet{HELP2021}, respectively.

\subsection{The Planck List of High-\textit{z} Source Candidates}
\label{PHz_catalogue}

\begin{figure*}
\includegraphics[width=\textwidth]{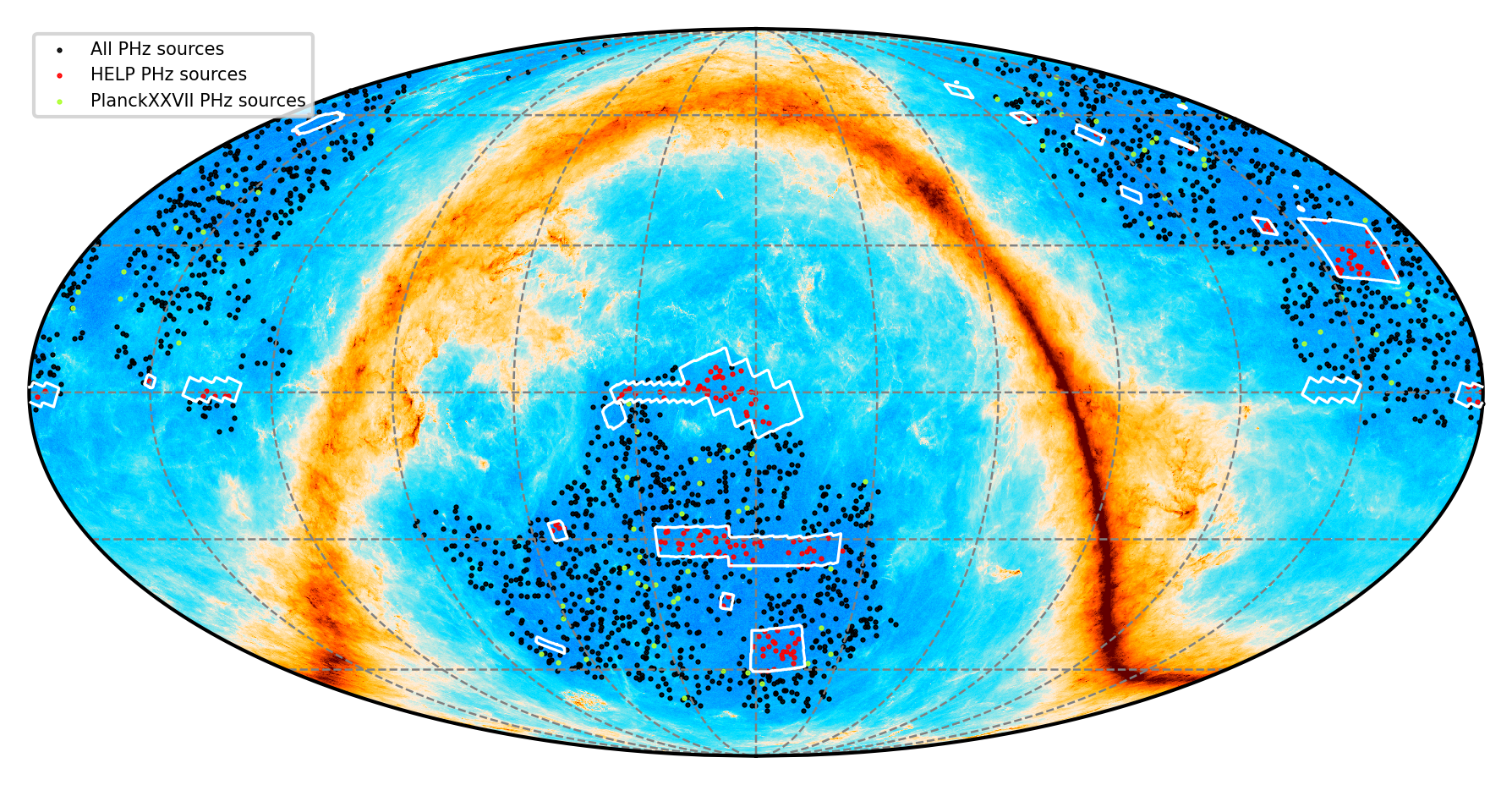}
\caption{Mollweide projection of the {\it Planck\/} 545-GHz map in equatorial coordinates with the HELP fields highlighted in white. PHz sources that were followed up in PlanckXXVII are shown in green (83 sources), the PHz sources that fall in the HELP fields are coloured red (187 sources) and all remaining PHz sources are shown in black (1874 sources).}
\label{HELP-fields}
\end{figure*}

{\it Planck\/} mapped the whole sky between 30 and 857\,GHz, including channels centred at 217, 353 and 545\,GHz \citep{PlanckI2013}. Using these maps, the Planck Collaboration generated a list of high-$z$ candidates by identifying red peaks in the CIB \citep{PlanckPHz2016}. Specifically, maps of the cleanest 25.8\,per cent of the sky (i.e., the least contaminated by Galactic emission) were used, and the \textit{Infrared Astronomical Satellite} ({\it IRAS\/}) 3-THz map was included in the analysis. The CMB was removed from each channel map using the 217-GHz map as a template, and Galactic cirrus was removed using the \textit{IRAS} 3-THz map. S/N$\,{>}\,$5 sources were identified in the 545-GHz `excess' map, constructed by subtracting a linear interpolation between the 353- and 857-GHz maps from the 545-GHz map. In addition, sources were required to have S/N$\,{>}\,$3 at 353 and 857\,GHz. Lastly, to avoid selecting cold Galactic clumps and radio galaxies, all sources with $S_{545}/S_{857}\,{<}\,0.5$ or $S_{353}/S_{545}\,{>}\,0.9$ were removed. The resulting catalogue consists of 2151 potential high-$z$ sources, spread across the northern and southern Galactic hemispheres (see Fig.~\ref{HELP-fields}). The focus on picking out the apparently coldest, potentially highest redshift objects makes the PHz catalogue complementary to the {\it Planck\/} compact source catalogues, with almost no overlap \citep{PlanckPHz2016}.

Across three proposal periods, a total of 228 {\it Planck}-identified sources were followed up with dedicated {\it Herschel}-SPIRE observations (PlanckXXVII): 204 sources were selected with an algorithm that resembled the final PHz selection process and the other 24 came from the PCCS. The 228-source sample was constructed based on preliminary and evolving {\it Planck\/} data, as new products were made available throughout the three SPIRE observation periods. Additionally, there was a bias for preferentially selecting bright sources with regular shapes to follow up with {\it Herschel}-SPIRE \citep{PlanckPHz2016}. As a result, many of the sources catalogued in PlanckXXVII did not end up in the final PHz list, and the sample was biased towards sources with high S/N, small angular sizes, small ellipticities and low extinctions when compared with the full PHz. Nonetheless, the selection process for the 228 sources was qualitatively similar to the PHz catalogue, requiring compact (at 5-arcmin resolution) source-detections at 545\,GHz.

Specifically, of the 228 {\it Planck}-identified sources presented in PlanckXXVII, 83/228 (36\,per cent) ended up in the final PHz catalogue. Due to these differences, the authors of \citet{PlanckPHz2016} remarked that `it is hard to draw definitive conclusions about the nature of the PHz sources based on the {\it Herschel\/} analysis'. In what follows, we study a large and unbiased sample of PHz sources using recently homogenized archival {\it Herschel}-SPIRE observations. For comparison, we repeated all of our analyses on the 228 PlanckXXVII sources (although in practice, we were unable to recover the {\it Herschel}-SPIRE maps of nine of the sources). Hereafter, the 228 {\it Planck}-identified sources followed up in PlanckXXVII will be referred to as the `PlanckXXVII sources'.

\subsection{The Herschel Extragalactic Legacy Project}
\label{HELP}

\begin{figure*}
\includegraphics[width=\textwidth]{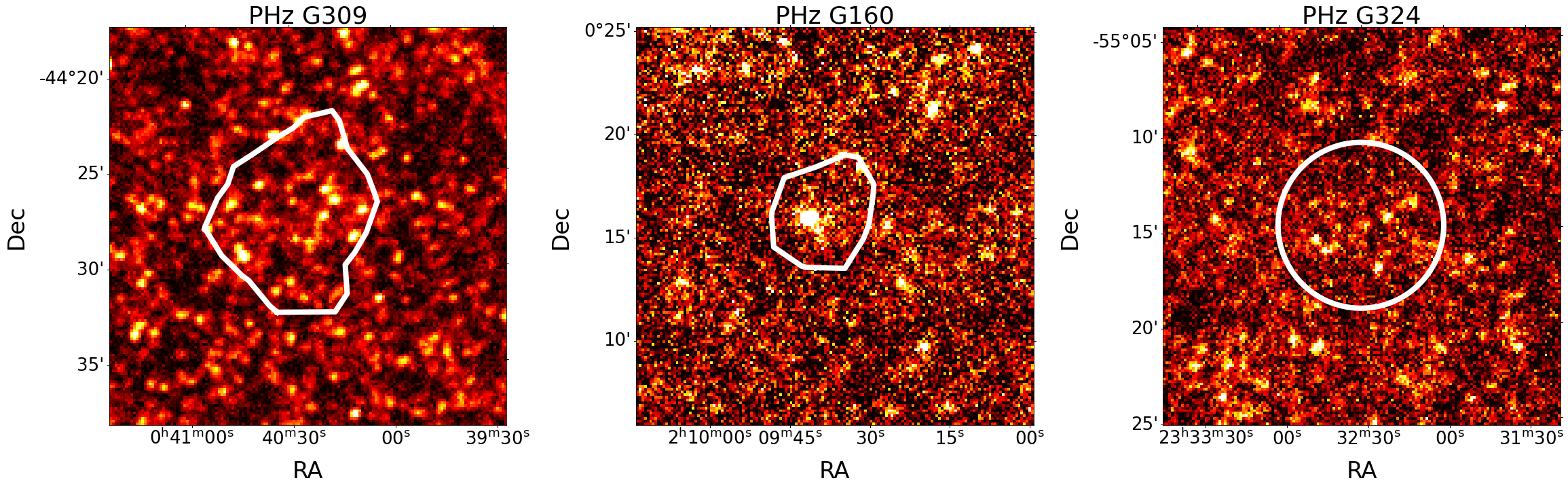}
\caption{{\it Herschel}-SPIRE 350-$\mu$m maps for a significantly overdense source (PHz G309.49$-$72.52), a source containing a known bright strong gravitationally-lensed galaxy (PHz G160.57$-$56.79) and a more typical source (PHz G324.51$-$58.45). The {\it Planck\/} 50\,per cent intensity IN regions are highlighted in white on G309 and G160, whereas a circular IN region is adopted for G324 because the {\it Planck\/} intensity contour does not close.}
\label{examplemaps}
\end{figure*}

\begin{table*}
\centering
\caption{Names, locations and areas of the HELP fields (from \citealt{HELP2021}), and their corresponding RMS noise levels at 250, 350 and 500\,$\mu$m. The total area on the sky of the HELP fields is 1270\,deg$^{2}$.}
\begin{tabular}{lcccccccc} 
 \hline
 Name & RA min & RA max & Dec min & Dec max & Area & 250\,$\mu$m (1$\,\sigma$) & 350\,$\mu$m (1$\,\sigma$) & 500\,$\mu$m (1$\,\sigma$)\\
 & [deg] & [deg] & [deg] & [deg] & [deg$^{2}$] & [mJy] & [mJy] & [mJy]\\
 \hline
  AKARI-NEP & 264.6 & 275.3 & \pp64.5 & \pp68.5 & \pz\pz9.2\pz & 11.0 & 10.1 & 11.3\\
  AKARI-SEP & \pz66.2 & \pz75.4 & $-$55.9 & $-$51.7 & \pz\pz8.7\pz & \pz7.3 & \pz8.1 & \pz9.1\\
  Bo\"otes & 215.7 & 220.6 & \pp32.2 & \pp36.1 & \pz11.4\pz & \pz6.2 & \pz7.1 & \pz8.0\\
  CDFS-SWIRE & \pz50.8 & \pz55.4 & $-$30.4 & $-$26.0 & \pz13.0\pz & \pz4.9 & \pz6.0 & \pz6.7\\
  COSMOS & 148.7 & 151.6 & \pz\pz0.8 & \pp\pz3.6 & \pz\pz\pz5.1\pz & \pz4.6 & \pz5.8 & \pz6.3\\
  EGS & 212.4 & 217.5 & \pp51.2 & \pp54.2 & \pz\pz3.6\pz & \pz5.3 & \pz6.4 & \pz7.2\\
  ELAIS-N1 & 237.9 & 247.9 & \pp52.4 & \pp57.5 & \pz13.5\pz & \pz6.5 & \pz7.3 & \pz8.3\\
  ELAIS-N2 & 246.1 & 252.3 & \pp39.1 & \pp43.0 & \pz\pz9.2\pz & \pz7.7 & \pz8.3 & \pz9.5\\
  ELAIS-S1 & \pz\pz6.4 & \pz11.2 & $-$45.5 & $-$41.6 & \pz\pz9.0\pz & \pz6.1 & \pz7.0 & \pz7.9\\
  GAMA-09 & 127.2 & 142.2 & \pz$-$2.5 & \pz\pp3.5 & \pz62.0\pz & 10.9 & 11.2 & 12.3\\
  GAMA-12 & 172.3 & 187.3 & \pz$-$3.5 & \pz\pp2.5 & \pz62.7\pz & 10.5 & 10.8 & 12.0\\
  GAMA-15 & 210.0 & 225.2 & \pz$-$2.5 & \pz\pp3.4 & \pz61.7\pz & 10.4 & 10.7 & 12.0\\
  HATLAS-NGP & 189.9 & 209.2 & \pp21.7 & \pp36.1 & 177.7\pz & 11.0 & 11.3 & 12.2\\
  HATLAS-SGP & 337.2 & \pz26.9 & $-$35.6 & $-$24.5 & 294.6\pz & 11.5 & 11.7 & 12.7\\
  HDF-N & 188.1 & 190.4 & \pp61.8 & \pp62.7 & \pz\pz0.67 & \pz3.6\ & \pz5.0 & \pz5.6\\
  Herschel-Stripe-82 & 348.4 & \pz36.2 & \pz$-$9.1 & \pz\pp8.9 & 363.4\pz & 15.3 & 15.4 & 16.7\\
  Lockman-SWIRE & 154.8 & 167.7 & \pp55.0 & \pp60.8 & \pz22.4\pz & \pz5.5 & \pz6.6 & \pz7.3\\
  SA13 & 197.6 & 198.5 & \pp42.4 & \pp43.0 & \pz\pz0.27 & \pz9.1 & \pz9.8 & 10.9\\
  SPIRE-NEP & 263.7 & 266.4 & \pp68.6 & \pp69.4 & \pz\pz0.6\pz & \pz4.0 & \pz5.0 & \pz5.5\\
  SSDF & 341.5 & \pz\pz2.1 & $-$60.5 & $-$48.5 & 110.4\pz & 10.9 & 11.1 & 12.7\\
  XMM-13hr & 202.9 & 204.4 & \pp37.4 & \pp38.5 & \pz\pz0.76 & \pz8.1 & \pz8.8 & \pz9.8\\
  XMM-LSS & \pz32.2 & \pz38.1 & \pz$-$7.5 & \pz$-$1.6 & \pz21.8\pz & \pz6.4 & \pz7.3 & \pz8.2\\
  xFLS & 255.6 & 262.5 & \pp57.9 & \pp60.8 & \pz\pz7.4\pz & \pz7.3 & \pz8.0 & \pz9.0\\
 \hline
\end{tabular}
\label{HELP-table}
\end{table*}

The {\it Herschel\/} satellite carried out a number of wide-field extragalactic surveys using the SPIRE instrument (see Fig.~\ref{HELP-fields} and Table~\ref{HELP-table}). HELP combines the {\it Herschel}-SPIRE survey fields into a consistent single data release spanning a total of 1270\,deg$^{2}$ on the sky \citep{HELP2021}. In particular, HELP DR1 provides homogeneous SPIRE 250-, 350- and 500-$\mu$m images for each of the HELP fields. The HELP fields avoid regions of bright dust emission from our own Galaxy, span a range of right ascension and include areas of sky near both ecliptic poles. As a result, there is significant overlap between the coverage of HELP fields and sources in the PHz catalogue, allowing us to select an unbiased sample of PHz sources at the SPIRE wavelengths.

A total of 194 sources from the PHz list fall within the HELP fields, six of which lie too close to the edge of the HELP field coverage and one falls within a region masked by HELP because of Galactic cirrus, resulting in a total of 187 usable sources. Hereafter, these 187 PHz sources that lie in the HELP fields will be referred to as the `PHz HELP sources'. By design of the follow up in PlanckXXVII, there is no overlap between the PHz HELP sources and the PlanckXXVII sources. However, the PHz HELP sample includes two sources that have previously been studied extensively, namely PHz G95.50$-$61.59 \citep{Flores-CachoG952016} and PHz G237.0$+$42.5 \citep{KoyamaG2372021, PollettaG2372021}.

\section{Observations and photometry}
\label{observations}

\subsection{SPIRE-detected galaxies within PHz sources}
\label{inoutgalaxies}
In order to study the number counts of SPIRE-detected galaxies in the PHz catalogue, we define an `IN region' in the same way as in PlanckXXVII.\footnote{To avoid confusion, we will reserve the word `source' for the {\it Planck}-selected targets and refer to the SPIRE-detected objects as `galaxies'.} We draw contours enclosing 50\,per cent of the {\it Planck\/} flux density at 545\,GHz (the frequency used to select PHz sources), and use the galaxies within this region for further analysis. Other follow-up studies have adopted a circular IN region centred on the {\it Planck\/} position \citep[e.g.,][]{GreensladeHerschelPCCS2018}. We find that using a contour on the {\it Planck\/} 545-GHz map generally encloses sources more effectively than a circular IN region. That is, when the galaxies making up a PHz source cover an extended region, {\it Planck\/} partially resolves the source and the contour is slightly larger than the {\it Planck\/} beam, whereas for individual bright galaxies, the contour is approximately the size of the beam (see Fig.~\ref{examplemaps}). As a result, defining the IN region using a {\it Planck\/} contour misses fewer contributing galaxies than a circular IN region (although the differences are not dramatic). For comparison, the contours have a median area of 59.2\,arcmin$^{2}$ on the sky, corresponding to a search radius of 4.34\,arcmin, which is slightly smaller than the 4.63\,arcmin radius adopted in \citet{GreensladeHerschelPCCS2018}. We determine the IN region contour by smoothing the 545-GHz map with a 2.0\,arcmin full width at half-maximum (FWHM) Gaussian, then find the contour on the smoothed 545-GHz map corresponding to 50\,per cent of the peak flux density of the source. For sources where the contour does not close or extends far beyond the {\it Herschel}-SPIRE map coverage, a circular IN region centred on the {\it Planck\/} position with a radius of 4.34\,arcmin is adopted, corresponding to the median contour area.

\subsection{Independent photometry}
\label{photometry}
Before performing photometry, we measure the noise at each wavelength across all of the HELP fields in a similar fashion to PlanckXXVII (Table~\ref{HELP-table}). Specifically, in each HELP field, for each SPIRE band, we construct histograms including all of the pixels in the map. The histograms have a mean of 0\,mJy, and appear Gaussian in shape, except for a positive tail of bright pixels corresponding to detected galaxies. We fit a Gaussian to each histogram in order to calculate the standard deviation of the pixel value distribution, $\sigma$, which quantifies the noise of the map (i.e., the combined confusion and detector noise). Note that all pixels, including those in the positive tail, are included when performing the Gaussian fit, thereby accounting for confusion noise (if instead the fit is restricted to the negative half of the histogram, the RMS noise levels change only slightly). For consistency with previous studies \citep[e.g., PlanckXXVII;][]{GreensladeHerschelPCCS2018}, we use the same flux density thresholds for galaxy detection across all HELP fields. Because the noise levels of the HELP fields vary (Table~\ref{HELP-table}), we adopt 250, 350 and 500\,$\mu$m flux density thresholds that are significantly larger than the noise level of all the fields (${\gtrsim}\,3\,\sigma$). Specifically, we choose flux density cutoffs of 4$\bar{\sigma}$ at 250, 350 and 500\,$\mu$m, that is, 4 times the mean noise of all the HELP fields. The cutoff flux densities are 32.0, 34.6 and 38.5\,mJy at 250, 350 and 500\,$\mu$m, respectively. These thresholds are comparable to those adopted in PlanckXXVII, although they are higher than the levels used in \citet{GreensladeHerschelPCCS2018}, where 25.4\,mJy was used at all three wavebands. We find that our results are not sensitive to the choice of flux density thresholds.

We detect {\it Herschel}-SPIRE galaxies independently in the 250-, 350- and 500-$\mu$m maps before matching galaxies across the three bands. Photometry is performed on sky-subtracted mean-0 maps using Gaussian point-spread functions (PSFs) with FWHMs of 18.15, 25.15 and 36.30\,arcsec for the 250-, 350- and 500-$\mu$m maps, respectively \citep{Griffin2010}. First, we convolve the signal map with the corresponding Gaussian PSF. Then, we search for peaks in the convolved signal map that fall within the IN region. All local peaks in the 250-, 350- and 500-$\mu$m convolved maps above the corresponding galaxy detection thresholds are independently identified as potential galaxies. The values and positions of peak pixels in the convolved maps are used as initial guesses for the galaxy fluxes and positions at the wavelength of detection. A Gaussian PSF fit is then performed on each detected galaxy in the non-convolved map, with a fixed FWHM (depending on the wavelength), and initial position/amplitude guesses from the convolved map. During PSF fitting, the PSF amplitude is allowed total freedom, but the positions are allowed to shift only slightly from the initial position ($0.5\,{\times}\,$FWHM in each direction) to avoid accidentally fitting the PSF to a separate nearby galaxy instead of the intended galaxy. Finally, the amplitudes and positions of the fitted Gaussian PSFs are added to the 250/350/500\,$\mu$m galaxy catalogue. 

When galaxy blending is present, which is especially common in the overdense sources discovered with {\it Planck}, a more careful fitting procedure is required. Whenever multiple galaxies are detected within two FWHMs of each other (corresponding to 6.0, 5.1 and 5.2 pixels at 250, 350 and 500\,$\mu$m, respectively), galaxy flux densities and positions are measured with simultaneous PSF fitting. That is, Gaussian PSFs are simultaneously fit to each of the nearby galaxies in the non-convolved map, avoiding the risk of inadvertently including the fluxes from other nearby galaxies during PSF fitting. In practice, twice the FWHM is a conservative threshold; galaxies separated by ${\gtrsim}\,2\,{\times}\,$FWHM are too far away for significant blending to occur, in which case simultaneously fitting the PSFs is equivalent to independently fitting Gaussian PSFs to both galaxies. After simultaneous PSF fitting, the amplitudes and positions of the fitted Gaussian PSFs are added to the 250/350/500\,$\mu$m galaxy catalogue. We find that our galaxy detection and photometry values are largely consistent with the results from other comparable codes, including codes developed specifically for {\it Herschel}-SPIRE observations (e.g., {\sc sussextractor}; \citealt{PearsonHIPE2014}).

\subsection{Galaxy matching}
\label{galaxymatching}
Galaxies are matched across bands in order to measure single-galaxy properties (e.g., colours), as well as to identify candidate strong gravitationally-lensed galaxies. Matching galaxies detected across the 250-, 350- and 500-$\mu$m maps is done by focusing on the galaxies detected at 350\,$\mu$m (also see PlanckXXVII). The 350-$\mu$m map has the advantage of higher angular resolution and lower noise than the 500-$\mu$m map, while also avoiding many of the low-$z$ galaxies that appear in the 250-$\mu$m map. Our band-merging procedure proceeds as follows. For each galaxy detected in the 350-$\mu$m map, if there is a galaxy within $0.5\,{\times}\,{\rm FWHM}$ of the 350-$\mu$m galaxy position in either the 250-$\mu$m map (FWHM 18.15\,arcsec) or the 500-$\mu$m map (FWHM 36.30\,arcsec), then the independently-calculated flux density is adopted from the corresponding map. If a galaxy is not detected within $0.5\,{\times}\,{\rm FWHM}$ in the 250- or 500-$\mu$m maps, we perform a `forced' Gaussian PSF fit in the 250- or 500-$\mu$m maps (for which there are no enforced thresholds). That is, we use the 350-$\mu$m galaxy position as the initial guess position and the value in the nearest pixel in the convolved 250- or 500-$\mu$m map as the initial guess flux in order to fit a Gaussian PSF (as in Section~\ref{photometry}). If multiple galaxies are detected within $0.5\,{\times}\,{\rm FWHM}$ of the 350-$\mu$m position at either 250 or 500\,$\mu$m, we instead perform simultaneous forced Gaussian PSF fitting at the positions of the close galaxies in the other two maps (i.e., if a galaxy detected at 350\,$\mu$m breaks up into multiple galaxies at 250\,$\mu$m, simultaneous PSF fitting is carried out at the positions of the galaxies in the 350- and 500-$\mu$m maps). To be consistent with independent galaxy detection, when forced PSF fitting is performed, we carry out simultaneous PSF fits on all galaxies found within $2\times{\rm FWHM}$ of the galaxy being matched across maps. For instance, if a galaxy at 350\,$\mu$m has a single corresponding galaxy within $0.5\,{\times}\,{\rm FWHM}$ at 250\,$\mu$m, but there is a second galaxy $1.5\,{\times}\,{\rm FWHM}$ away at 250\,$\mu$m, simultaneous Gaussian PSF fitting on both of the galaxies is performed at 350\,$\mu$m and 500\,$\mu$m, to avoid biases due to galaxy blending. If a 350-$\mu$m galaxy is successfully matched to a single galaxy at 250\,$\mu$m, or is revealed to break up into multiple galaxies, we record the positions of the galaxy at 250\,$\mu$m because of the superior angular resolution of the 250-$\mu$m map (otherwise, the 350-$\mu$m position is used). With this procedure, we generate a catalogue of galaxies with flux densities measured at 250, 350 and 500\,$\mu$m.

\section{Source statistics}
\label{sourcestats}

\subsection{\textit{Planck} flux densities}
\label{planckfluxes}
\begin{figure*}
\includegraphics[width=\textwidth]{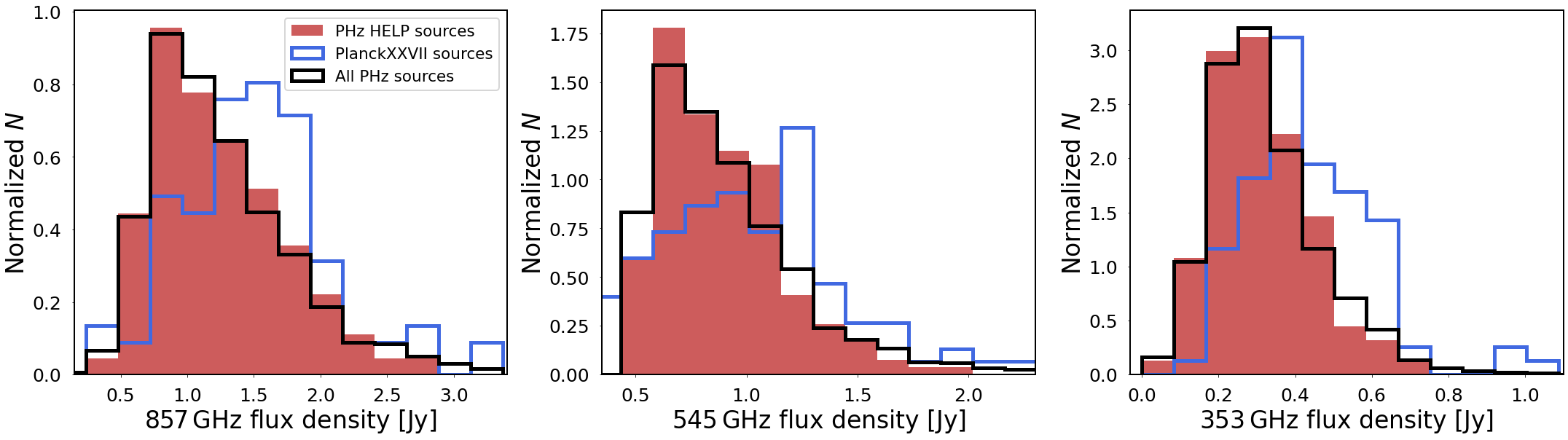}
\caption{Histograms of {\it Planck\/} flux densities for PHz HELP sources, the PlanckXXVII sources (those with {\it Planck\/} reported fluxes) and all PHz sources. The distribution of {\it Planck\/} flux densities for the PHz HELP sources is statistically indistinguishable from the entire PHz sample, whereas the PlanckXXVII sources are generally brighter at 857, 545 and 353\,GHz.}
\label{planckfluxdists}
\end{figure*}

As described in Section~\ref{PHz_catalogue}, the PHz sources were selected using {\it Planck\/} flux densities, or more specifically, the `excess' flux density at 545\,GHz. PlanckXXVII followed up a sample of 228 bright sources selected in a similar fashion as the PHz catologue, of which 83 ended up in the final version of the PHz catalogue. To evaluate the statistical differences resulting from this selection process, in Fig.~\ref{planckfluxdists} we show the distributions of the {\it Planck\/} flux densities for the 187 PHz HELP sources, about 100 of the PlanckXXVII sources (the sources for which {\it Planck\/} fluxes are publicly available) and all 2151 PHz sources. PlanckXXVII sources that came from earlier, non-public, PHz catalogue-like selection processes are necessarily excluded. The two-sample Kolmogorov–Smirnov (KS) test rejects the null hypothesis that the PlanckXXVII sources are drawn from the same sample as the full PHz catalogue with $p$-values of $9.7 \times 10^{-8}$, $1.6 \times 10^{-4}$ and $2.7 \times 10^{-10}$ at 857, 545 and 353\,GHz, respectively. For the PHz HELP sources, on the other hand, the two-sample KS test gives a $p$-value of $0.87$, $0.69$ and $0.90$ at 857, 545 and 353\,GHz, respectively. In other words, the 187 PHz HELP sources are drawn from the same distribution as the entire PHz sample, and are therefore representative of the overall PHz catalogue. These trends are similar across other PHz source properties, such as FWHMs and ellipticities. It is worth noting that differences between the PlanckXXVII source flux densities and the PHz source flux densities (as well as other properties) are not simply caused by the inclusion of PCCS sources in the PlanckXXVII sample. We see an equally significant difference when we restrict the comparison to the 83 PlanckXXVII sources that ended up in the final PHz catalogue, indicating that there were biases in the selection of PlanckXXVII sources (see \citealt{PlanckPHz2016} for a discussion of these selection effects).

\subsection{Number counts}
\label{numcounts}
\begin{figure*}
\includegraphics[width=\textwidth]{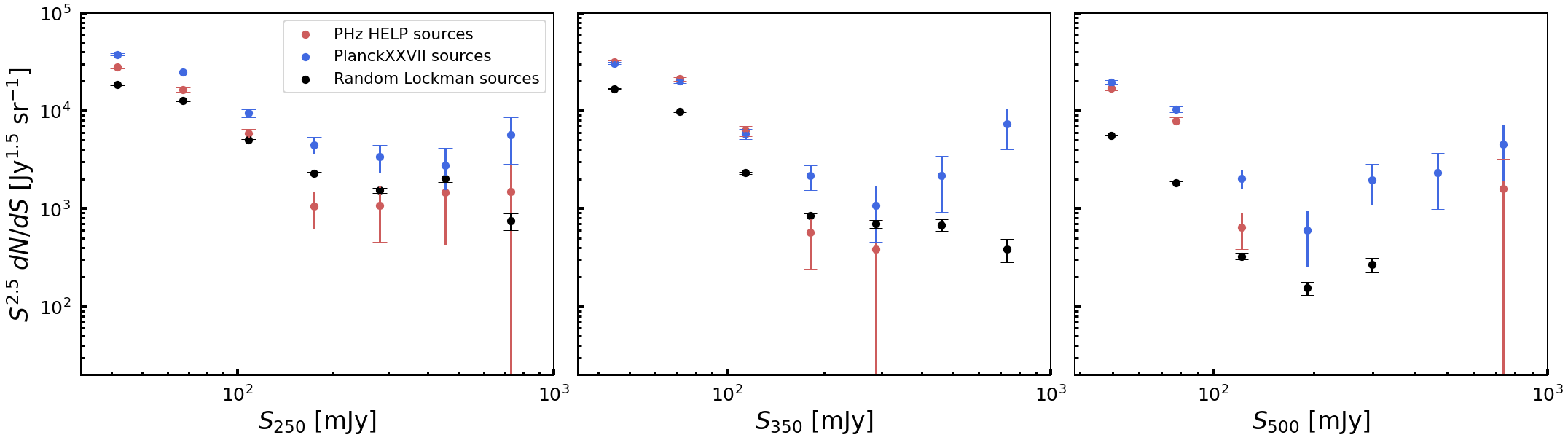}
\caption{Differential Euclidean-normalized number counts for galaxies within our sample of PHz HELP sources (red), galaxies in the PlanckXXVII sources (blue) and 10{,}000 random positions in the Lockman-SWIRE field (black). Number counts are intended to be compared relative to one another and have no corrections applied. Both the PHz HELP sources and the PlanckXXVII sources have excess number counts when compared with random Lockman-SWIRE positions, but the PHz HELP sources have lower 250-$\mu$m number counts than the PlanckXXVII sources, as well as lower number counts above 100\,mJy across all three bands.}
\label{num_counts}
\end{figure*}

In Fig.~\ref{num_counts} we show Euclidean-normalized differential number counts of SPIRE-detected galaxies, $S^{2.5}\,dN/dS$, where $N$ is the total number of galaxies per steradian in the IN regions and $S$ is the flux density at 250, 350 or 500\,$\mu$m. Number counts are calculated using the galaxies detected independently in each of the three bands and are intended to be compared relative to one another, so no corrections (such as for completeness) have been applied. As in PlanckXXVII, we use the Lockman-SWIRE field to compare between the PHz HELP and PlanckXXVII sources. To compare the number counts of IN region galaxies with the number counts of random locations, we compute the number counts at 10{,}000 random positions in Lockman-SWIRE within a circle of radius 4.34\,arcmin (the median size of the IN regions).

Similarly to the PlanckXXVII sources, when compared with random locations on the sky, the PHz HELP sources have a significant excess in number counts below approximately 100\,mJy across the 250-, 350- and 500-$\mu$m bands, with the most substantial differences seen at 500\,$\mu$m. This confirms that typical PHz sources contain an exceptional number of galaxies, just as the preferentially-selected PlanckXXVII sources do. That being said, the PHz HELP sources have lower number counts at 250\,$\mu$m than the PlanckXXVII sources. Additionally, above about 100\,mJy, there are substantial differences between the PHz HELP and PlanckXXVII number counts. As found in PlanckXXVII, the PlanckXXVII sources contain significantly higher numbers of SPIRE galaxies at large flux densities compared to random positions, the brightest of which are now known to be strong gravitational lenses. Across all three bands, the HELP sources have lower numbers of SPIRE galaxies at large flux densities compared to the PlanckXXVII sample, potentially indicating that fewer of the PHz HELP sources contain strong gravitationally-lensed galaxies. This is investigated further in Section~\ref{PC_GLcandidates}.

\subsection{Projected galaxy densities}
\label{overdensities}
\begin{figure*}
\includegraphics[width=\textwidth]{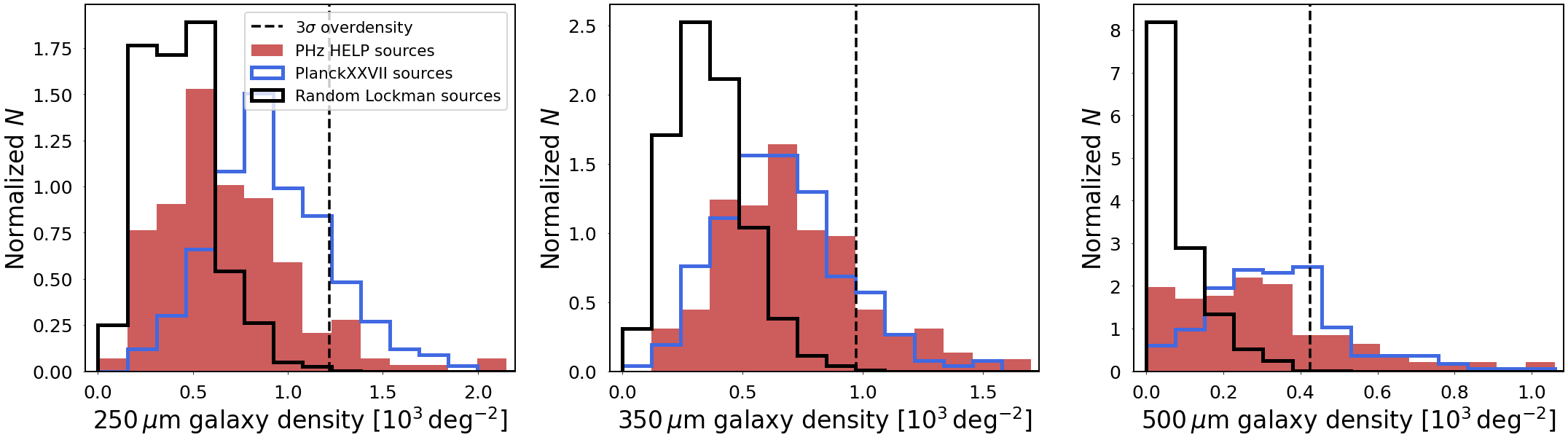}
\caption{Histograms of IN region galaxy densities for the PHz HELP sources (red), the PlanckXXVII sources (blue) and 10{,}000 random positions in the Lockman-SWIRE field (black), with $3\,\sigma$ overdensity thresholds shown as vertical lines.The PHz HELP sources and the PlanckXXVII sources are overdense when compared with random positions at 250, 350 and 500\,$\mu$m, with the largest difference seen at 500\,$\mu$m. The PlanckXXVII sources are significantly more overdense than the PHz HELP sources at 250\,$\mu$m, and the PHz HELP sources have a slightly greater positive tail at 350\,$\mu$m.}
\label{sourcedens}
\end{figure*}

We next quantify the projected overdensities of our PHz HELP sources. In Fig.~\ref{sourcedens} we compare the number densities of IN region galaxies for the PHz HELP sources, the PlanckXXVII sources and 10{,}000 random positions in the Lockman-SWIRE field. Just as we found in Section~\ref{numcounts}, the PHz HELP sources are significantly overdense when compared with random positions, across all three bands. The PlanckXXVII sources appear more overdense at 250\,$\mu$m compared to the PHz HELP sources, whereas galaxy densisites appear more comparable at 350 and 500\,$\mu$m. 

\begin{figure*}
\includegraphics[width=\textwidth]{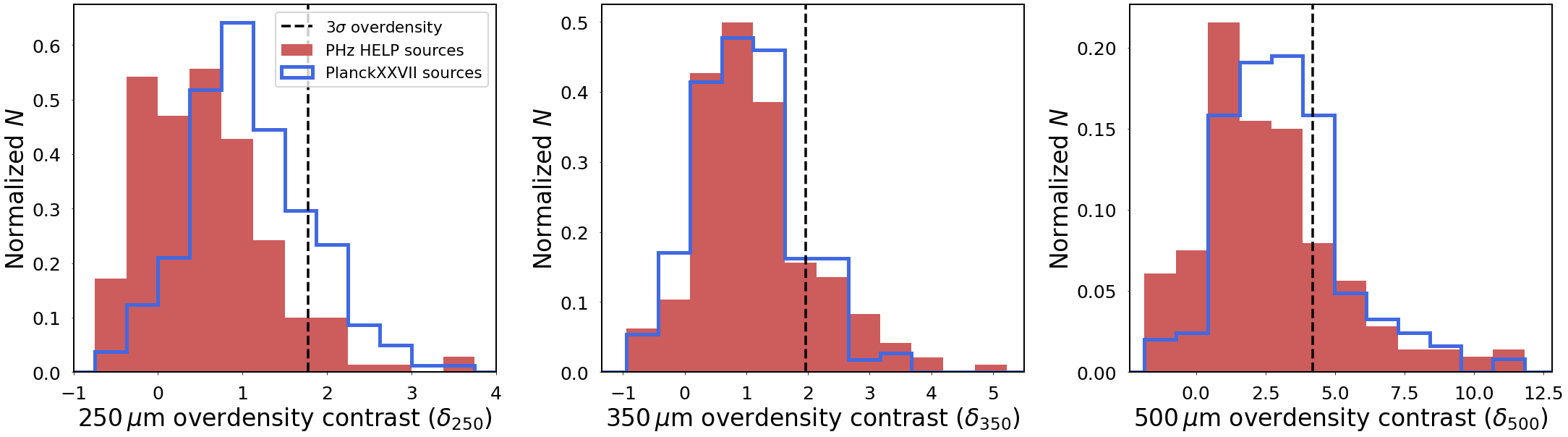}
\caption{Histograms of IN galaxy overdensity contrasts for the PHz HELP sources (red) and the PlanckXXVII sources (blue), with the $3\,\sigma$ overdensity thresholds from Fig.~\ref{sourcedens} shown as vertical lines. The PHz HELP sources are significantly less overdense than the PlanckXXVII sources at 250\,$\mu$m and there is a slightly larger positive tail of overdense 350-$\mu$m PHz HELP sources.}
\label{overcontrasts}
\end{figure*}

In order to take into account the background densities, as was done in PlanckXXVII, we define the overdensity contrast of a {\it Planck\/} source as
\begin{equation} 
\label{overdens_contrast}
\delta_{\lambda} = \frac{\rho_{\mathrm{gal}} - \Bar{\rho}_{\mathrm{rand}}}{\Bar{\rho}_{\mathrm{rand}}},
\end{equation}
\noindent
where $\rho_{\mathrm{gal}}$ is the galaxy density within the contour of the PHz HELP/PlanckXXVII source and $\Bar{\rho}_{\mathrm{rand}}$ is the mean galaxy density of the 10{,}000 random positions in Lockman-SWIRE. In Fig.~\ref{overcontrasts} we show the overdensity contrasts across all three SPIRE bands for our sample of PHz HELP sources and for the PlanckXXVII sources. As suggested by Fig.~\ref{num_counts}, the overdensity contrast distribution at 250\,$\mu$m for the PlanckXXVII sources is shifted to the right when compared to the PHz HELP sources. At 350\,$\mu$m and 500\,$\mu$m, the galaxy densities are comparable, although the PHz HELP sources have a slightly greater positive tail at 350\,$\mu$m and more sources with low galaxy density at 500\,$\mu$m. The two-sample KS test confidently rejects (at ${>}\,8\,\sigma$) the null-hypothesis that the PHz HELP source galaxy densities are drawn from the same distribution as the random location galaxy densities across 250, 350 and 500\,$\mu$m. Again, this confirms the exceptional nature of typical PHz sources, demonstrating that on average the PHz sources are similarly overdense at 350 and 500\,$\mu$m to the preferentially-selected PlanckXXVII sources.

\subsection{Colours}
\label{colours}
\begin{figure*}
\includegraphics[width=0.9\textwidth]{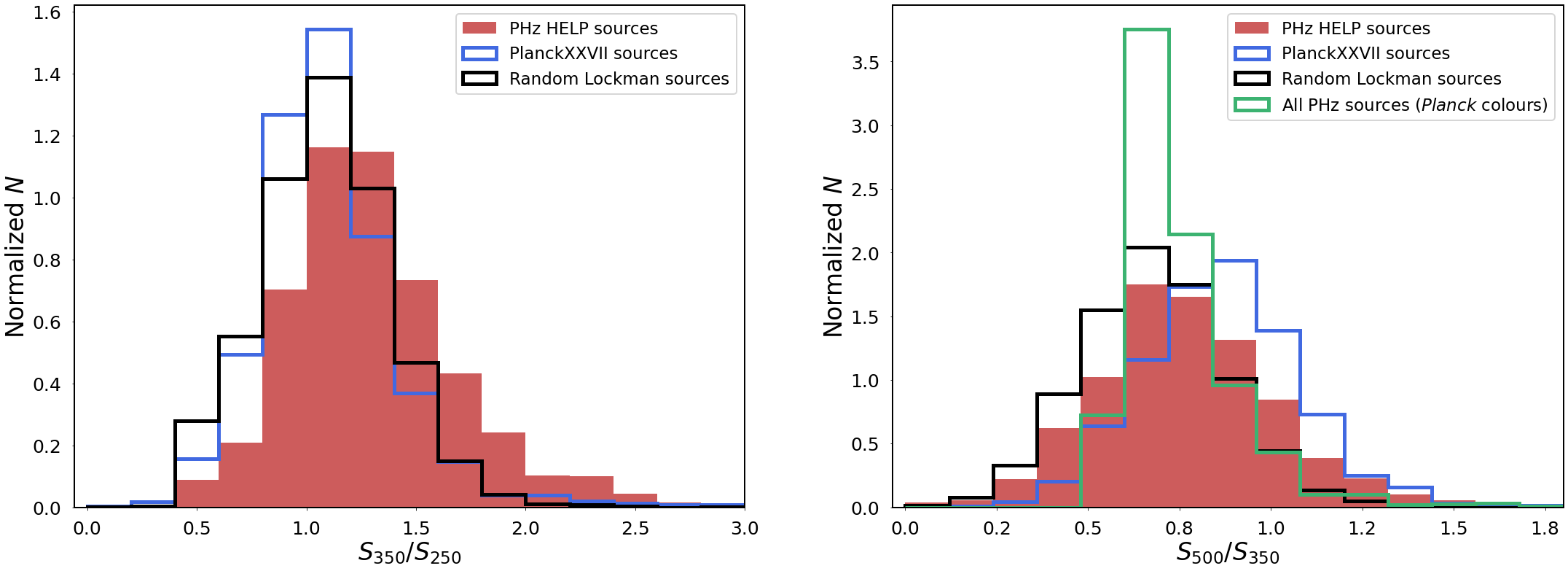}
\caption{Histograms of the ratio $S_{350}/S_{250}$ (left) and $S_{500}/S_{350}$ (right) for galaxies in the PHz HELP sources (red), the PlanckXXVII sources (blue) and 10{,}000 random positions in the Lockman-SWIRE field (black). {\it Planck\/} $S_{500}/S_{350}$ colours are shown in green (unfortunately, {\it Planck\/} does not have a 250-$\mu$m channel for comparison of $S_{350}/S_{250}$). The PHz HELP sources have a tail of red galaxies with large $S_{350}/S_{250}$ and $S_{500}/S_{350}$ values, whereas the PlanckXXVII sources have a tail of galaxies with large $S_{500}/S_{350}$ colours, but no excess of galaxies with large $S_{350}/S_{250}$ colours.}
\label{colourdists}
\end{figure*}

After matching galaxies across the 250-, 350- and 500-$\mu$m bands, colours are quantified by comparing flux density ratios. Figure~\ref{colourdists} shows the distributions of $S_{350}/S_{250}$ and $S_{500}/S_{350}$ for all of the galaxies within in the PHz HELP sources, the PlanckXXVII sources and 10{,}000 random Lockman-SWIRE positions. The PHz HELP sources display a clear excess of $S_{350}/S_{250}$ and $S_{500}/S_{350}$ ratios compared to random positions, indicative of an excess of red galaxies. The PlanckXXVII sources display a more significant excess of $S_{500}/S_{350}$ ratios compared to random positions; however, the distribution of $S_{350}/S_{250}$ is more consistent with random positions (these distributions are comparable to Fig.~7 from PlanckXXVII). Quantitatively, when PHz HELP colours are compared to random sources, we find that the null hypothesis is rejected at 20.3$\,\sigma$ and 14.2$\,\sigma$ for the $S_{350}/S_{250}$ colours and the $S_{500}/S_{350}$ colours, respectively. These differences demonstrate that the average PHz source contains an excess of red (and potentially higher $z$) sources, as was previously concluded for the preferentially-selected PlanckXXVII sources. However, we also find significant differences between the PHz HELP and PlanckXXVII galaxy colours. The two-sample KS test rejects (at 16.7$\,\sigma$) the null hypothesis that the distribution of $S_{350}/S_{250}$ colours for the PHz HELP sources is drawn from the same distribution as the PlanckXXVII sources. Similarly, for the distribution of $S_{500}/S_{350}$ colours, the KS test rejects the null hypothesis at 12.6$\,\sigma$. Furthermore, the peak in PHz HELP $S_{500}/S_{350}$ colours better aligns with the peak for all PHz sources from {\it Planck\/} colours.

\begin{figure*}
\includegraphics[width=\textwidth]{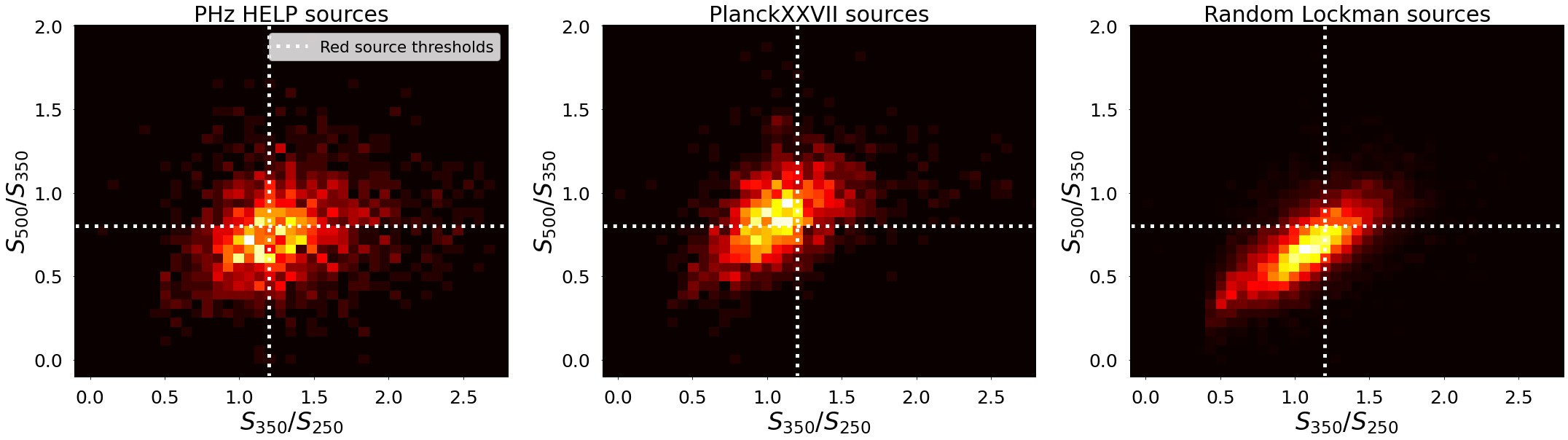}
\caption{2D histograms of $S_{350}/S_{250}$ versus $S_{500}/S_{350}$ for galaxies from the PHz HELP sources (left), the PlanckXXVII sources (middle) and 10{,}000 random positions in the Lockman-SWIRE field (right). White dotted lines show red galaxy thresholds, where $S_{350}/S_{250}\,{=}\,1.2$ and $S_{500}/S_{350}\,{=}\,0.8$. Galaxies detected in the PHz HELP sources, PlanckXXVII sources and random positions differ both in histogram location and number of galaxies with large $S_{350}/S_{250}$ colours and $S_{500}/S_{350}$ colours.}
\label{colourscatter}
\end{figure*}

\begin{figure*}
\includegraphics[width=0.9\textwidth]{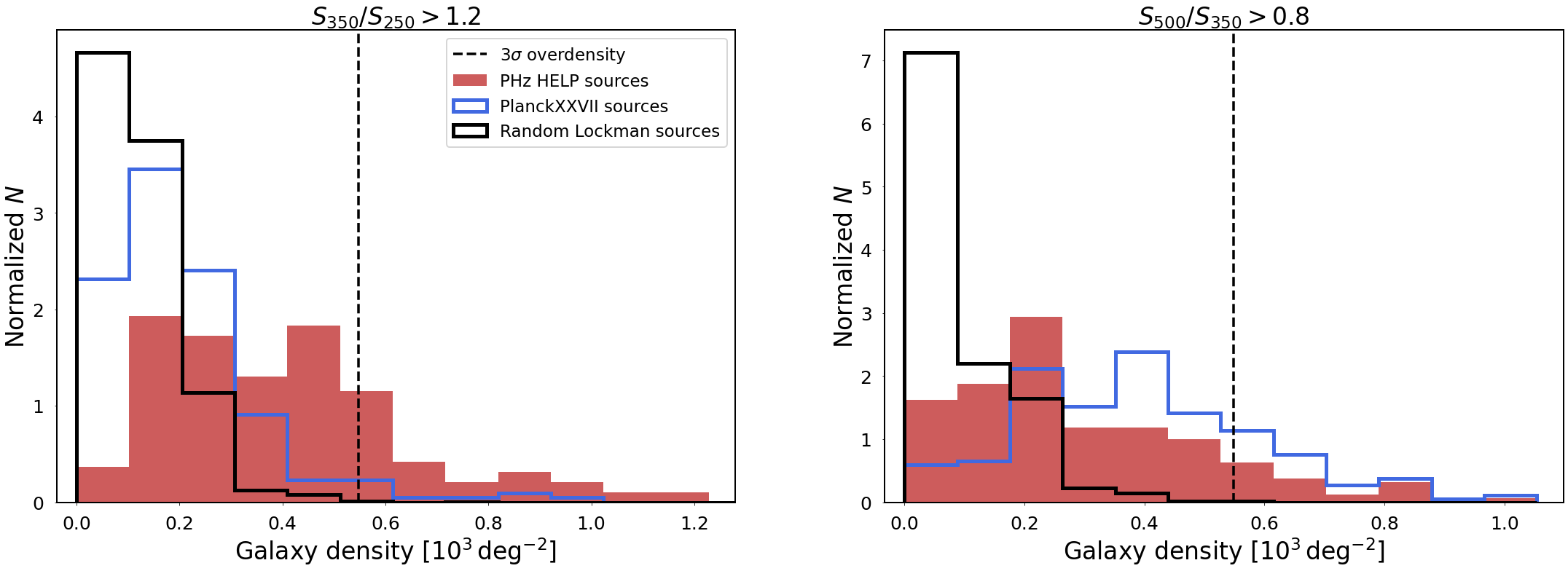}
\caption{Histograms of the density of red galaxies (left for $S_{350}/S_{250}\,{>}\,$1.2, right for $S_{500}/S_{350}\,{>}\,$0.8) for the PHz HELP sources (red), the PlanckXXVII sources (blue) and 10{,}000 random positions (black), with $3\,\sigma$ red galaxy overdensity thresholds shown as vertical lines. The PHz HELP sources have significantly higher densities of red galaxies with $S_{350}/S_{250}\,{>}\,$1.2, whereas the PlanckXXVII sources have higher densities of red galaxies with $S_{500}/S_{350}\,{>}\,$0.8.}
\label{redsourcehists}
\end{figure*}

To further illustrate the differences in galaxy colours between the PHz HELP, PlanckXXVII and random Lockman-SWIRE positions, Fig.~\ref{colourscatter} shows 2D histograms of $S_{350}/S_{250}$ versus $S_{500}/S_{350}$. The PHz HELP sources have a clear tail of red galaxies extending to the right, which is not seen in the PlanckXXVII sources or the random sources. However, the PlanckXXVII colour distribution appears shifted upwards when compared to the other two distributions. Based on the distribution of random galaxy colours, we set the thresholds $S_{350}/S_{250}\,{=}\,1.2$ and $S_{500}/S_{350}\,{=}\,0.8$ to select the reddest galaxies, falling in the upper-right region of Fig.~\ref{colourscatter}. Overlaying these thresholds on the 2D colour histograms of the PHz HELP and PlanckXXVII sources illustrates the differences between the colour distributions; when compared with the PlanckXXVII sources, significantly more PHz HELP sources lie above the $S_{350}/S_{250}\,{=}\,1.2$ threshold and noticeably fewer lie above the $S_{500}/S_{350}\,{=}\,0.8$ threshold. Density histograms of red galaxies that fall above these two cuts confirm these trends (see Fig.~\ref{redsourcehists}). The two-sample KS test rejects the null hypothesis that the distribution of red PHz HELP sources is the same as the distribution the red PlanckXXVII sources at 8.0$\,\sigma$ for the $S_{350}/S_{250}\,{>}\,1.2$ sources and 5.3$\,\sigma$ for the $S_{500}/S_{350}\,{>}\,0.8$ sources.

Based on the 2D colour distributions, we can also infer that fewer PHz HELP galaxies peak in flux density at 250\,$\mu$m as compared with both random galaxies and PlanckXXVII galaxies. Quantitatively, of random galaxies detected in the Lockman-SWIRE field, a total of 38\,per cent peak at 250\,$\mu$m, and comparatively, 37\,per cent of PlanckXXVII galaxies peak at 250\,$\mu$m. For the PHz HELP sources, on the other hand, we find that only 21\,per cent of galaxies peak at 250\,$\mu$m. Instead, most PHz HELP galaxies (66\,per cent) peak at 350\,$\mu$m, whereas only 41\,per cent of galaxies in the PlanckXXVII sources peak at 350\,$\mu$m. In this way, the PHz HELP sources typically consist of redder galaxies than the PlanckXXVII sources do. This suggests that the distribution of galaxy redshifts in PHz HELP sources may differ from the distribution of redshifts in PlanckXXVII sources, a possibility that is further explored in Section~\ref{MBB fits}; however, it is difficult to draw conclusions about galaxy redshifts with SPIRE photometry alone. It is also worth noting that PlanckXXVII reports that none of the brightest galaxies within the PlanckXXVII sources peak at 250\,$\mu$m (i.e., within each PlanckXXVII source, the brightest galaxy across all three bands is brightest at 350/500\,$\mu$m), while we find that a significant fraction of the brightest PlanckXXVII galaxies peak at 250\,$\mu$m (56\,per cent), and a smaller, but still significant, fraction of the brightest PHz HELP galaxies peak at 250\,$\mu$m (37\,per cent).

\subsection{Photometric redshifts and star-formation rates}
\label{MBB fits}
Owing to the fact that the rest-frame far-IR emission observed by SPIRE probes a relatively simple, thermal portion of a galaxy's spectral energy distribution (SED), SPIRE colours can be used to provide crude estimates of photometric redshifts, luminosities and therefore SFRs, after applying various simplifying assumptions \citep[e.g.,][]{blain2003,draine2003}. Assuming that any AGN contribution is sub-dominant, the SPIRE bands are sensitive to thermal dust emission generated by star formation, characterized by a modified blackbody distribution with a single dust temperature. However, due to redshifting, we observe a dust temperature that is decreased by a factor of $1\,{+}\,z$, preventing us from determining both the intrinsic dust temperature and the redshift simultaneously. Nonetheless, if we assume a typical dust temperature motivated by studies of similar SPIRE-detected galaxies, we can obtain rough redshift estimates and SFRs, which should at least be useful for comparitive studies.

\begin{figure*}
\includegraphics[width=0.9\textwidth]{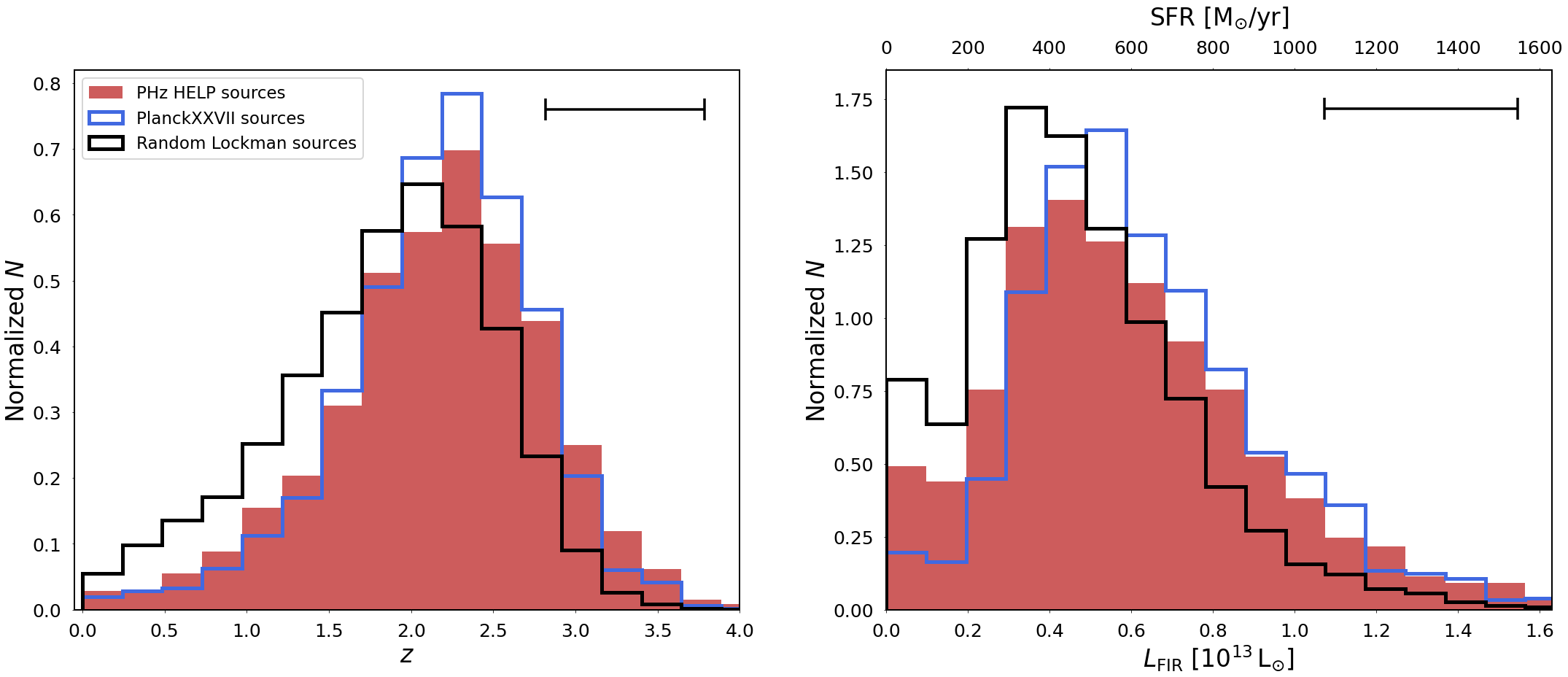}
\caption{Histograms of SPIRE photometric redshifts (left) and far-IR luminosities/SFRs (right) for galaxies in the PHz HELP sources (red), the PlanckXXVII sources (blue) and 10{,}000 random positions in the Lockman-SWIRE field (black). See Section~\ref{MBB fits} for details on the estimation of redshifts and far-IR luminosities/SFRs. The PHz HELP and PlanckXXVII sources have a positive tail of high-$z$ galaxies when compared with random positions. Although the PHz HELP sources generally have larger far-IR luminosities than random galaxies, they are generally smaller than the PlanckXXVII far-IR luminosities. When the adopted dust temperature is changed by $\pm 5$\,K, the histograms are shifted left/right as shown by the error bars in the top right, but the same trends persist.}
\label{zSFRdists}
\end{figure*}

For each galaxy in our sample, we fit the three SPIRE flux densities to a modified blackbody SED of the form
\begin{equation}\label{mbb}
    S_{\nu_{\rm obs}} \propto \left(\frac{\nu_{\rm obs}(1+z)}{\nu_0}\right)^{(\beta + 3)} \frac{1}{{\rm e}^{h \nu_{\rm obs} (1+z) / k_{\rm B} T} - 1},
\end{equation}
\noindent
where $\nu_{\rm obs}$ is the observed frequency, $z$ is the redshift, $\nu_0$ is an arbitrary normalization constant fixed to 3000\,GHz, $h$ is the Planck constant, $k_{\rm B}$ is the Boltzman constant, $\beta$ is the dust emissivity index, which is fixed to 1.5 (the same value adopted in PlanckXXVII) and $T$ is the dust temperature. In order to make progress, we set $T\,{=}\,30\,$K for all of the galaxies in the analysis. This is in line with the typical dust temperatures measured in large samples of $z\,{=}\,1$--2 submm-bright galaxies \citep[e.g.,][]{roseboom2012,magnelli2014}. Of course in reality, galaxies exhibit a wide range of temperatures, but we can nonetheless compare results between the PHz HELP sources, the PlanckXXVII sources and random positions. In Fig.~\ref{zSFRdists} (left panel) we show the resulting distribution of photometric redshifts for our PHz HELP sources, compared to the distribution for the PlanckXXVII sources and 10,000 random Lockman positions. We find similar trends when instead adopting a dust temperature of $T\,{=}\,25\,$K or $T\,{=}\,35\,$K, and use the change in the median of the redshift distribution to make an estimate of the redshift uncertainty of $\delta z \approx 1$.

Next, we calculate far-IR luminosities ($L_{\rm FIR}$) by integrating our fits to Eq.~\ref{mbb} between 8 and 1000\,$\mu$m (in the rest-frame), and convert these to SFRs using a conversion factor of $0.95\,{\times}\,10^{-10}\,{\rm M}_{\odot}\,{\rm yr}^{-1}\,{\rm L}_{\odot}^{-1}$, which assumes a Chabrier initial mass function \citep{chabrier2003}. The distributions of $L_{\rm FIR}$ and SFR are shown in Fig.~\ref{zSFRdists} (right panel). We again find similar trends when the assumed dust temperature is changed by $\pm 5$\,K, giving an uncertainty of $\delta L_{\rm FIR} \approx 5 \times 10^{12}\,{\rm L}_{\odot}$ and $\delta \mathrm{SFR} \approx 5 \times 10^{2}\,{\rm M}_{\odot}\,{\rm yr}^{-1}$.

In agreement with PlanckXXVII, we find that the redshift distribution of galaxies in the PHz HELP sources peaks at $z\,{\approx}\,2$. The PHz HELP galaxy redshifts are largely comparable to the PlanckXXVII galaxy redshifts and both have a positive tail of high-$z$ galaxies when compared with galaxies found in random positions. The KS test confidently rejects the null hypothesis (at ${>}\,15\,\sigma$) that the PHz HELP redshifts are drawn from the same distribution as the random galaxy redshifts. 

It remains unknown whether the PHz sources are predominantly protoclusters of physically-associated galaxies at the same redshift or line-of-sight alignments of multiple groups of galaxies \citep[e.g.,][]{Negrello2017, Gouin2022}, as discussed further in Sect.~\ref{discussion}. By comparing the SPIRE photometric redshifts of multiple galaxies detected within the same PHz source, we can attempt to quantify the presence of line-of-sight alignments within the PHz HELP sources, but we find that the SPIRE photometric redshifts are too imprecise ($\delta z \approx 1$) to distinguish between overdensities at a single redshift from random alignments along the line-of-sight.

Both the PHz HELP sources and PlanckXXVII sources generally have higher far-IR luminosities/SFRs than random sources (the KS test puts the significance of the differences at ${>}\,15\,\sigma$). Furthermore, the far-IR luminosity/SFR distribution of the PlanckXXVII sources is shifted to the right of the distribution of the PHz HELP sources; the difference between the PHz HELP and PlanckXXVII sources is significant at ${>}\,12\,\sigma$ by the KS test. The larger far-IR luminosities/SFRs of galaxies within the PlanckXXVII sources agrees with the larger flux densities of the PlanckXXVII sources, which is seen in the galaxy number counts (Fig.~\ref{num_counts}). Nonetheless, in the more representative PHz HELP sources, we still find an excess of high far-IR/SFR galaxies when compared with random positions.

\subsection{Source stacking and surface brightness profiles}
\label{radialfluxes}

\begin{figure*}
\includegraphics[width=\textwidth]{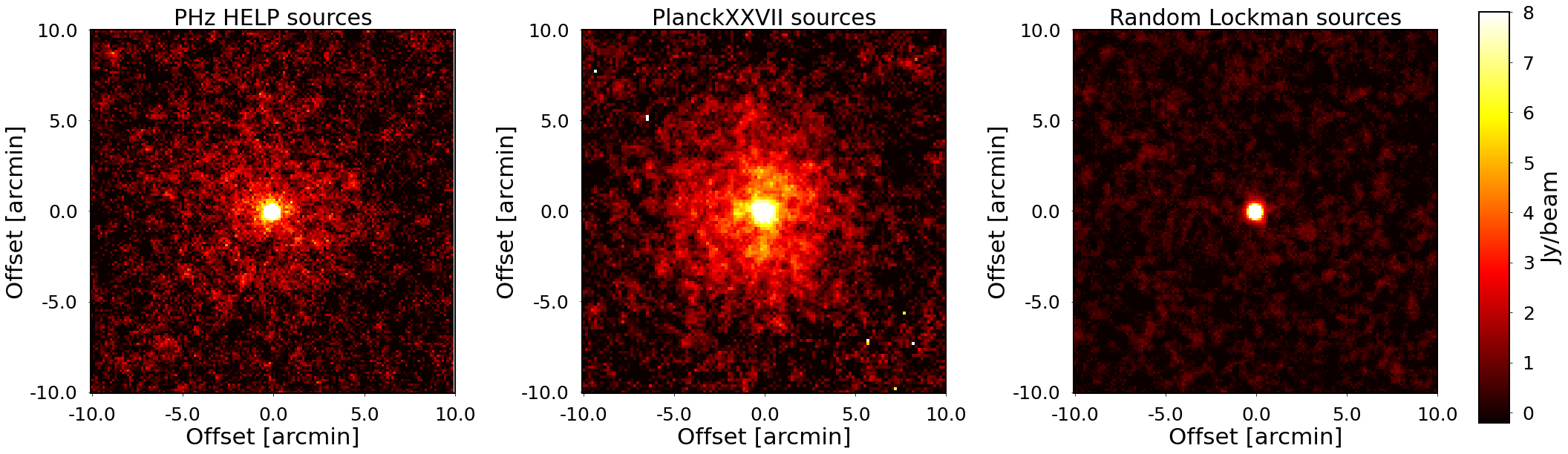}
\caption{Mean stacked 350-$\mu$m maps for the PHz HELP sources (left), the PlanckXXVII sources (middle) and random positions (right). Here 20\,arcmin$\,{\times}\,20\,$arcmin cutouts were made around each source, centred on the brightest galaxy detected at 350\,$\mu$m in the given source, and averaged together after subtracting the background. Since the PlanckXXVII sources are on average brighter (having been selected for their high S/N values), the resulting stack appears brighter than the PHz HELP stack.}
\label{stackplots}
\end{figure*}

\begin{figure*}
\includegraphics[width=0.9\textwidth]{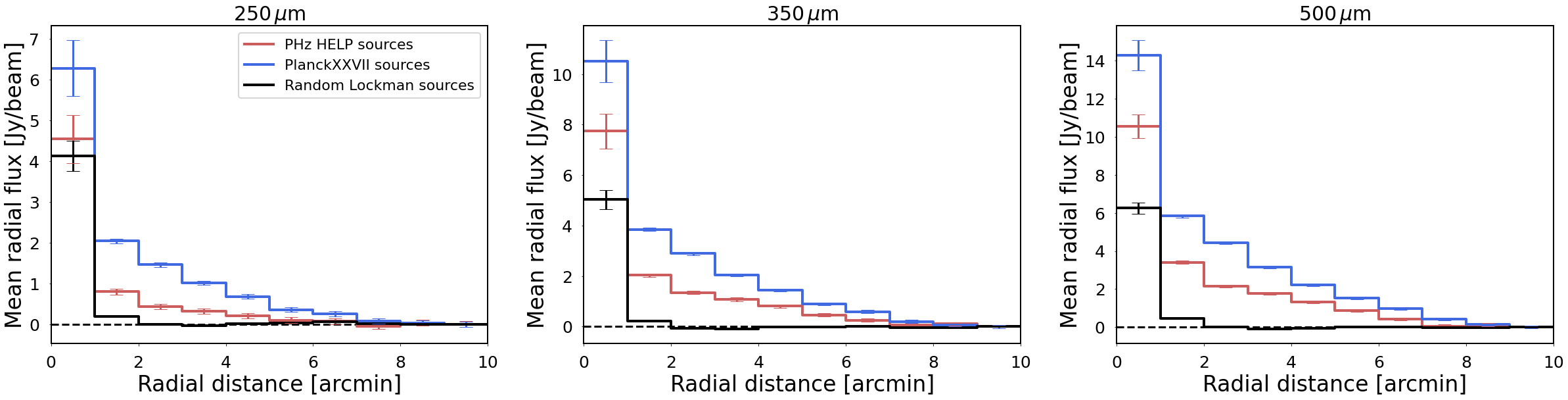}
\caption{Radial surface brightness distributions of the stacked 250-, 350- and 500-$\mu$m maps for the PHz HELP sources (red), PlanckXXVII sources (blue) and random positions in the Lockman-SWIRE field (black). Confirming the visual trends seen in the stacked 350-$\mu$m maps (Fig.~\ref{stackplots}), the PHz HELP stacks are less bright than the PlanckXXVII stacks across 250, 350 and 500\,$\mu$m. Nonetheless, the PHz HELP stacks have significant extended structure when compared with random galaxies, which are effectively point sources.}
\label{radialplots}
\end{figure*}

\begin{figure*}
\includegraphics[width=0.9\textwidth]{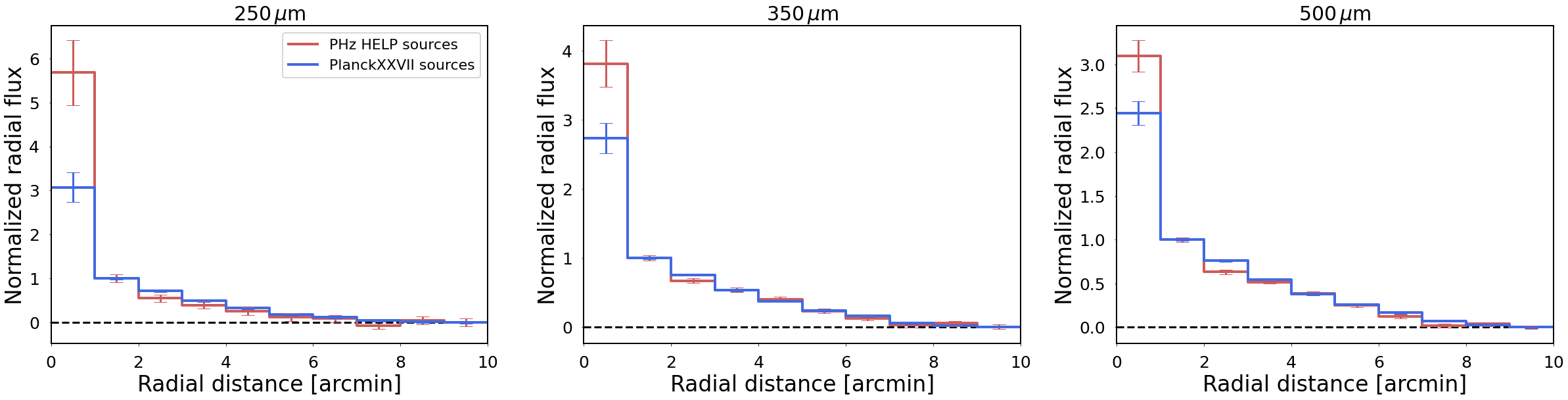}
\caption{Same as Fig.~\ref{radialplots}, but normalized to 1.0\,Jy\,beam$^{-1}$ in the 1.0--2.0\,arcmin bin. Although the PlanckXXVII sources are on average brighter than the PHz HELP sources, we find that their normalized radial profiles match closely, except in the central $1.0$\,arcmin bin, wherein the PHz HELP stacks are comparatively brighter.}
\label{radialplotsnormalized}
\end{figure*}

We next investigate the spatial properties of the PHz HELP sources through a stacking analysis. Because we are primarily interested in the spatial distributions of the overdense sources, we remove the sources that consist of a single bright galaxy (namely the {\it Planck} GEMS from the PlanckXXVII sample, and the bright gravitational lens G160 from the PHz HELP sample). We create mean image stacks by averaging the 250-, 350- and 500-$\mu$m maps centred on the position of the brightest 350-$\mu$m galaxy in each IN region.\footnote{We explored other stacking choices as well, including stacking on the flux-weighted centroid positions, and found similar trends.} Before computing the average, we set the background of each image to 0\,Jy\,beam$^{-1}$ by subtracting the mean intensity value in pixels that are ${>}\,9$\,arcmin from the centre. In Fig.~\ref{stackplots} we show the resulting stacks at 350\,$\mu$m for the PHz HELP sources, PlanckXXVII sources and random positions in the Lockman-SWIRE field centred on the brightest galaxy in each 4.34\,arcmin radius circular region. It is clear from the 350-$\mu$m stacks that the PHz HELP sources are significantly less bright than the PlanckXXVII sources. Nonetheless, similar to the PlanckXXVII stack, the PHz HELP stack has clustering extending out to about 5\,arcmin from the centre that is not seen around random galaxies in the Lockman-SWIRE field. This confirms that average PHz sources display clustering, just as the preferentially-selected bright PlanckXXVII sources do.

To quantify the extended nature of the PHz HELP and PlanckXXVII stacks, we plot the radial surface brightness profiles of the PHz HELP, PlanckXXVII and random position stacks (see Fig.~\ref{radialplots}). Confirming the visual conclusions drawn from Fig.~\ref{stackplots}, the PHz HELP stacks are less bright than the PlanckXXVII stacks, but still clearly extended when compared with random sources across 250, 350 and 500\,$\mu$m. In comparison with the random sources, the PHz HELP sources display the strongest clustering at 500\,$\mu$m and the weakest at 250\,$\mu$m. This is likely caused by the PHz selection process, which relied primarily on detecting excess emission at 500 $\mu$m. To directly compare the PHz HELP and PlanckXXVII radial profiles, we arbitrarily normalize both profiles to 1.0\,Jy\,beam$^{-1}$ in the 1.0--2.0\,arcmin bin (Fig.~\ref{radialplotsnormalized}). The PHz HELP and PlanckXXVII radial profiles closely match outside the central 1.0\,arcmin bin; however, the PHz HELP sources are brighter within this central bin, suggesting that the PHz HELP sources are (relatively) more centrally-concentrated. 

\begin{figure*}
\includegraphics[width=0.9\textwidth]{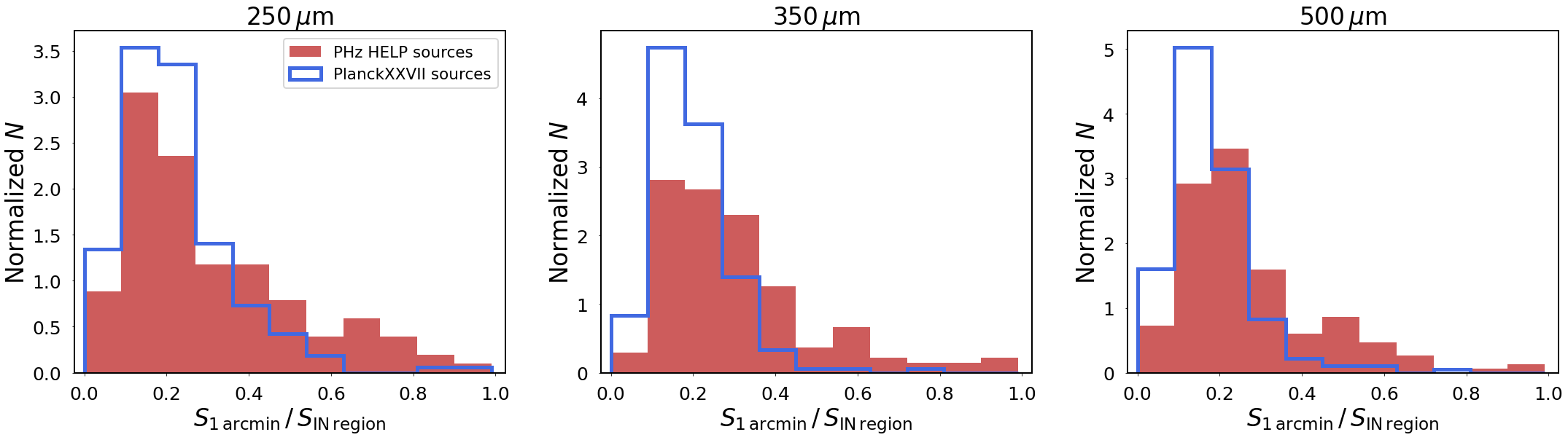}
\caption{Histograms of the ratio of the flux density contained within 1\,arcmin of the brightest 350-$\mu$m galaxy (used to centre the stacks) to the total flux density contained within the IN region ($S_{\mathrm{1\,arcmin}}/S_{\mathrm{IN\,region}}$). For this plot, we always use a 4.34\,arcmin radius circle for the IN region so that differences in $S_{\mathrm{1 arcmin}}/S_{\mathrm{IN\,region}}$ are not caused by the differences in IN region size. Across 250, 350 and 500\,$\mu$m, the PHz HELP sources have a larger $S_{\mathrm{1\,arcmin}}/S_{\mathrm{IN\,region}}$ value when compared with the PlanckXXVII sources, indicating that they are comparatively more centrally-concentrated.}
\label{fluxratiodists}
\end{figure*}

To confirm that the PHz HELP sources are genuinely more centrally-concentrated than the PlanckXXVII sources, we compare the ratios of the flux density contained within 1.0\,arcmin of the brightest galaxy to the total flux density of the IN region ($S_{\mathrm{1\,arcmin}}/S_{\mathrm{IN\,region}}$; see Fig.~\ref{fluxratiodists}). For the purpose of this plot, we adopt a 4.34\,arcmin radius circle for the IN region of each source so that differences in $S_{\mathrm{1\,arcmin}}/S_{\mathrm{IN\,region}}$ are not caused by variations in the IN region sizes. As alluded to by the normalized radial profiles, a greater fraction of the PHz HELP source flux densities come from the 1.0\,arcmin region around the brightest galaxy when compared to the PlanckXXVII sources. According to the KS test, the differences between the PHz HELP and PlanckXXVII sources is significant at ${>}\,3\,\sigma$ across the 250-, 350- and 500-$\mu$m distributions. We thus conclude that the PHz HELP sources are less bright, but more centrally-concentrated than the PlanckXXVII sources.

\section{Protocluster and strong gravitational lens candidates}
\label{PC_GLcandidates}
Since there are now a number of confirmed strong gravitational lenses \citep{CanamerasPlanckGEMS2015} and protoclusters \citep{Flores-CachoG952016,KneisslG732019, KoyamaG2372021} originating from {\it Planck} observations, we expect that similar objects should exist in our PHz HELP sample. Here we pursue a comprehensive search for protocluster and strong gravitational lens candidates, with the motivation to provide a representative list of targets for future surveys to further investigate the nature of the PHz sources, and to bolster the sample size of such objects, thereby enabling more detailed studies of star formation, early galaxy clustering and dark matter substructure.

\subsection{Protocluster candidates}
\label{PCcandidates}

Because the photometric redshift information offered by the three far-infrared SPIRE bands is too uncertain to identify redshift clustering about a single target, we turn to the spatial overdensity data to identify the most promising protocluster candidates. We then statistically search for clustering within the combined photometric redshift distribution of the best protocluster candidates.

\subsubsection{3$\,\sigma$ spatial overdensities}
\label{3sigmacuts}

To search for the most likely protocluster candidates in our sample of 187 PHz sources, we now estimate the significance of the observed spatial overdensities. Some studies (e.g., PlanckXXVII) have relied on kernel density estimator techniques to determine the statistical significance of spatial overdensities, which sometimes results in high significances for both overdense sources and random locations in the Lockman-SWIRE field. To obtain more easily-interpretable significance estimates, we instead turn to a comparison with the distributions of galaxy densities from random positions in the Lockman-SWIRE field, which allow us to estimate 3$\,\sigma$ overdensity thresholds (see e.g., \citealt{GreensladeHerschelPCCS2018, ChengSCUBA2019, ChengSCUBA2020}). In order to construct a homogeneous, representative sample of promising PHz protocluster candidates, we use the total galaxy densities in Figs.~\ref{sourcedens} and \ref{overcontrasts} (we separately explore the possibility of instead selecting protocluster candidates with red galaxy densities below). Using the distribution of galaxy densities from 10{,}000 random positions, we interpret the number of random sources with $N$ or more galaxies as a probability ($N/10{,}000$), thereby estimating the statistical significance of a source containing $N$ or more galaxies and subsequently the significance of the corresponding galaxy density. A total of 12 random sources contain 20 or more $S_{250}\,{>}\,32.0\,$mJy galaxies, corresponding to a probability of $12/10{,}000$, or 3.0$\,\sigma$. A random source made up of 20 galaxies has a surface density of $1220$\,deg$^{-2}$, providing an approximate 3$\,\sigma$ galaxy surface density threshold. Likewise, at 350\,$\mu$m, 3$\,\sigma$ random sources of $S_{350}\,{>}\,34.6\,$mJy galaxies contain 16 galaxies, corresponding to a 3$\,\sigma$ surface density of $970$\,deg$^{-2}$. Lastly, the 3$\,\sigma$ surface density of $S_{500}\,{>}\,38.5\,$mJy galaxies is $430$\,deg$^{-2}$. In Figs.~\ref{sourcedens} and \ref{overcontrasts} we show these resulting 3$\,\sigma$ thresholds as vertical dashed lines.

With these thresholds, we identify the most promising potential protoclusters as those that are ${>}\,3\,\sigma$ overdense at 250, 350 or 500\,$\mu$m (see Table~\ref{protoclustertable}). One downside of this approach is that some sources pass the $3\,\sigma$ overdensity threshold because they contain few (${\lesssim}\,$5) galaxies within a small {\it Planck\/} intensity contour (${\lesssim}\,$20\,arcmin$^{2}$) rather than containing many sources extended over several arcmin. To exclude these artificially-overdense sources from the list of candidate protoclusters, we adopt a conservative, arbitrary threshold of ${>}\,$5 galaxies at 350\,$\mu$m, which results in a total of 40 ${>}\,3\,\sigma$ protocluster candidates (changing the exact minimum galaxy threshold would include or exclude only a few sources).

In total, these 40 promising protocluster candidates make up 21\,per cent of the total PHz HELP sample, including nine sources that are ${>}\,3\,\sigma$ overdense at 250\,$\mu$m (4.8\,per cent), 26 at 350\,$\mu$m (14\,per cent) and 36 at 500\,$\mu$m (19\,per cent). When this analysis is repeated on the PlanckXXVII sources, we find that 57/228 (25\,per cent) are ${>}\,3\,\sigma$ overdense, of which 28 are ${>}\,3\,\sigma$ overdense at 250\,$\mu$m (12\,per cent), 24 at 350\,$\mu$m (10\,per cent) and 41 at 500\,$\mu$m (18\,per cent). Although the PHz HELP sources are significantly less bright than the preferentially-selected PlanckXXVII sources (see Section~\ref{radialfluxes}), there is a similar fraction of ${>}\,3\,\sigma$ overdensities in the PHz HELP and PlanckXXVII sources. Additionally, a smaller fraction of the PHz HELP sources are overdense at 250\,$\mu$m and a larger fraction are overdense at 350\,$\mu$m, consistent with the observed differences in galaxy colours.

For comparison with the PCCS, \citet{GreensladeHerschelPCCS2018} found a total of 27 ${>}\,3\,\sigma$ overdense protocluster candidates out of 354 PCCS sources (8\,per cent; many of these sources are overdense at 250\,$\mu$m), a substantially lower fraction, supporting the uniqueness of the PHz catalogue in containing many overdensities of red galaxies. However, because the PCCS contains significantly more sources than the PHz catalogue, the density on the sky of ${>}\,3\,\sigma$ overdense protocluster candidates in the PCCS catalogue ($3.3\,{\times}\,10^{-2}$\,deg$^{-2}$) is comparable to the density in the PHz catalogue ($3.1\,{\times}\,10^{-2}$\,deg$^{-2}$). It is worth noting that the PHz HELP sources that do not pass the 3$\,\sigma$ overdensity threshold are still interesting; the stacking analysis in Section~\ref{radialfluxes} demonstrated that typical PHz sources are still significantly more extended than random locations on the sky. In fact, neither of the two previously-studied PHz HELP sources, PHz G95.50$-$61.59 \citep{Flores-CachoG952016} and PHz G237.0$+$42.5 \citep{KoyamaG2372021, PollettaG2372021}, are $3\,\sigma$ overdense in the SPIRE maps with our galaxy detection thresholds.

To identify the most promising high-$z$ protocluster candidates in the PHz HELP sources, we could instead search for the sources that contain the most red galaxies (Fig.~\ref{redsourcehists}). Repeating the process of identifying 3$\,\sigma$ thresholds using the distribution of densities from the random Lockman sources, the 3$\,\sigma$ red galaxy density thresholds are shown as vertical lines in Fig.~\ref{redsourcehists}. After searching for PHz sources that lie above the 3$\,\sigma$ threshold of $S_{350}/S_{250}\,{>}\,$1.2 galaxies or the 3$\,\sigma$ threshold of $S_{500}/S_{350}\,{>}\,$0.8 galaxies (and requiring ${>}\,$5 galaxies at 350\,$\mu$m as before), we find 38 candidate protoclusters, a similar fraction to above. In addition to these criteria identifying a similar number of candidate protoclusters, there is significant overlap between the 40 candidates from above and the 38 red candidates; 27 of the 38 red candidates (71\,per cent) are included among the 40 overdense locations, supporting the potential high-$z$ nature of the protocluster candidates. For completeness, we report the 11 lower-significance red protocluster candidates in Table~\ref{protoclustertable}; however, we focus on the 40 homogeneously-selected sources as our main protocluster candidates.

\subsubsection{Photometric redshift clustering in 3$\,\sigma$ spatial overdensities}

Although SPIRE photometric redshifts for single galaxies are imprecise ($\delta z \approx 1$), we may still attempt to search for photometric redshift clustering by statistically averaging over the 40 best protocluster sources. To do so, we follow a simplified version of an algorithm detailed by \citet{Castignani2014}, which utilizes a background photometric redshift distribution to calculate Poisson probabilities within redshift bins. We define the background distribution (i.e., the number of galaxies per square degree per redshift) to be the distribution we derive from 10{,}000 random positions in the Lockman-SWIRE field (Fig.~\ref{zSFRdists}).

Next, we compare the photometric redshift distributions in each protocluster source with the background distribution. The redshift interval from 0--5 is divided into small redshift bins, allowing us to count the number of galaxies with photometric redshifts within a width $\Delta z$ of each bin. If the background number density of galaxies within the same redshift bin (say, bin $i$) is $\langle n_i \rangle$, and the solid angle of the protocluster candidate source is $\Omega$, then the probability of obtaining $N_i$ or more galaxies in the absence of clustering is given by the Poisson distribution,
\begin{equation}
    P_{\rm Poiss,i} = \sum_{k=N_i}^{\infty} \frac{(\langle n_i \rangle \Omega)^k}{k!}e^{- \langle n_i \rangle \Omega}.
\end{equation}
\noindent
Thus, the null hypothesis of no clustering is rejected with probability
\begin{equation}\label{clustering_eq}
    P_{\rm clust,i} = 1 - P_{\rm Poiss,i} = \sum_{k=0}^{N_i-1} \frac{(\langle n_i \rangle \Omega)^k}{k!}e^{- \langle n_i \rangle \Omega}
\end{equation}

\begin{figure}
\includegraphics[width=0.45\textwidth]{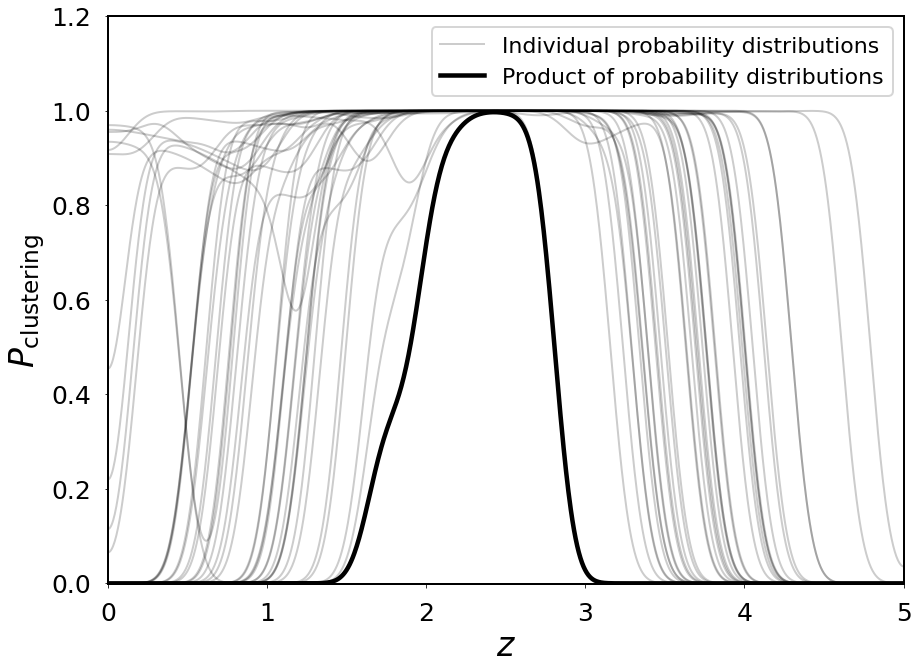}
\caption{Smoothed photometric redshift clustering probability distribution for the 40 ${>}\,3\,\sigma$ overdense protocluster candidates using redshifts bins of $\Delta z\,{=}\,1.0$ (see Eq.~\ref{clustering_eq}). We calculate the clustering probability for each of the 40 sources individually (faint grey lines), and then multiply the probability distributions to obtain the mean distribution across all sources (solid black line). The peak between $z\,{=}\,2$ and 3 indicates that the spatial overdensities in these 40 sources are not spread out uniformly in photometric redshift space, but are instead predominately found within this redshift range.}
\label{zclustering}
\end{figure}

The results from this calculation depend on the chosen redshift bins, which should be comparable to the photometric redshift uncertainty. We therefore fix $\Delta z\,{=}\,1$, and note that \citet{Castignani2014} provides a more generalized framework for incorporating arbitrary redshift bin sizes, but we do not expect a more detailed calculation to provide much more information because our photometric redshift uncertainties are large ($\delta z \approx 1$). Thus, for each redshift bin in each of the 40 protocluster candidate sources, we calculate $P_{\rm clust}$ using Eq.~\ref{clustering_eq} with $\Delta z\,{=}\,1$, and show the results in Fig.~\ref{zclustering} (faint grey lines). Note that the results here have been smoothed by a Gaussian kernel with a width of 0.1, which is much smaller than the photometric redshift uncertainty, and therefore should not introduce any biases.

Next, to quantify the photometric redshift clustering probability of the full protocluster candidate sample, we compute the product of the individual probability distributions (Fig.~\ref{zclustering}, solid black line). This operation effectively averages the likelihoods across the sample, leaving us with the mean probability of clustering over all 40 best protocluster candidates (see Fig.~\ref{zclustering}). The final probability distribution shows a significant peak around $z\,{=}\,2$--3, which indicates that the spatial overdensities in these 40 sources are not spread out uniformly in photometric redshift space, but are instead predominately found within this redshift range; repeating this analysis on the 38 red protocluster candidates produces a very similar probability distribution. However, we cannot say whether these sources are dominated by line-of-sight alignments within this redshift range; solving this issue will require spectroscopic follow-up observations.

\subsection{Strong gravitational lens candidates}
\label{MLlensed}
In PlanckXXVII, {\it Herschel}-SPIRE follow-up observations of 228 {\it Planck}-selected sources led to the discovery of 11 of the brightest known gravitationally-lensed submm galaxies (`{\it Planck}'s dusty GEMS'; \citealt{CanamerasPlanckGEMS2015}). Due to the exceptional brightness of the GEMS and the relative paucity of other SPIRE galaxies in these {\it Planck\/} sources, identification of such bright gravitational lenses turned out not to be very challenging. PlanckXXVII classified sources as either overdensities or candidate gravitationally-lensed galaxies using two quantities: the flux density of the brightest galaxy in the source (at any wavelength); and the wavelength at which the brightest galaxy peaks. Specifically, any sources containing a galaxy with flux density ${>}\,400$\,mJy were identified as candidate lens systems and all other sources whose brightest galaxy peaks at 350 or 500\,$\mu$m were identified as overdensities. This classification scheme identified seven of the PlanckXXVII sources as candidate lensed galaxies, and the authors selected an additional five candidate lensed galaxies by visual inspection (all of which have peak flux ${\gtrsim}\,300$\,mJy and appear as the only bright galaxy in the source). These 12 sources consisted of one previously-discovered strong gravitationally-lensed galaxy and 11 new lensing candidates, which have since been confirmed as bright gravitationally-lensed galaxies making up the GEMS sample \citep{CanamerasPlanckGEMS2015}.

\begin{figure}
\includegraphics[width=0.42\textwidth]{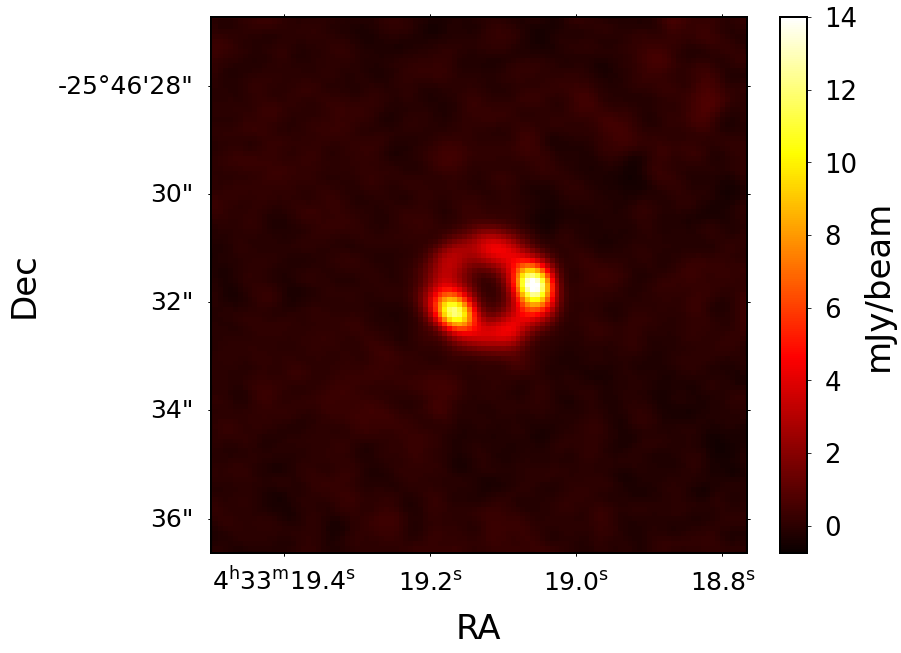}
\caption{ALMA image at 870\,$\mu$m of the candidate gravitationally-lensed PlanckXXVII source G224. The obvious Einstein ring confirms G224 as a new {\it Planck}-identified gravitationally-lensed galaxy.  It is somewhat dimmer than the ultra-bright {\it Planck\/} dusty GEMS, but still ${\sim}\,200\,$mJy at 500\,$\mu$m. ALMA observed G224 during Cycle 7 (PID 2019.1.01155.S, PI R. Hill).}
\label{G224ALMA}
\end{figure}

Only one PHz HELP source, G160, contains a single exceptionally bright SPIRE galaxy, like the GEMs, and turns out to be a previously-discovered strong gravitationally-lensed galaxy \citep{GeachG1602015, HarringtonG1602016}. However, not all strong gravitationally-lensed galaxies are as bright as {\it Planck}'s dusty GEMS; the PHz sources could contain high-$z$ strong gravitationally-lensed galaxies of somewhat lower brightness, comparable to other known gravitational lenses \citep[e.g.,][]{NegrelloLenses2010,WeissLensed2013}. Furthermore, unlike the GEMS, overdense PHz sources may contain high-$z$ strong gravitationally-lensed galaxies, making them difficult to identify visually. Studies of {\it Herschel\/} wide-field surveys have used a simple detection threshold of 100\,mJy at 500\,$\mu$m to identify candidate lensed galaxies. Applying this criterion to the galaxies within the PHz HELP sources, we find a total of just four sufficiently bright galaxies. To search for more candidates, here we use a machine-learning (ML) approach to select gravitationally-lensed galaxies in the PHz HELP sources.
 
\subsubsection{Feature selection}
The choice of features to input into an ML model plays a crucial role in successful classification \cite[see, for example][]{Liu2019}, in this case the identification of gravitationally-lensed galaxies. Due to the small training sample size and the unresolved nature of high-$z$ galaxies in SPIRE maps, we find that convolutional neural networks trained on images of sources struggle to successfully distinguish between lensed and non-lensed galaxies. Instead of providing images as inputs to the ML model, we choose to train an ML model using information about individual galaxies. This has the advantage of avoiding biases in the classification of galaxies based on other nearby galaxies (unlike visual identification), as well as compressing the total amount of information fed to the ML model. Furthermore, physically, we expect the brightnesses and colours to be the main pieces of information used to identify candidate gravitationally-lensed galaxies in these lower-resolution images \citep{Coppin2006, Negrello2007}. As a result, we provide the flux densities of individual IN region galaxies at 250, 350 and 500\,$\mu$m as inputs to the ML model. With the three galaxy flux densities as the input, a simple single-layer neural network (NN) classifier consistently trains to 100\,per cent accuracy on the test set (see below).

\subsubsection{Training and test sets}
\label{traindata}
In contrast to PlanckXXVII, we look for gravitationally-lensed galaxies by classifying the individual SPIRE galaxies in our {\it Planck\/} sources, as opposed to entire sources, allowing for the possibility of finding a gravitational lens within an overdensity. As such, we require a training set consisting of individual lensed galaxies (positive examples) and non-lensed galaxies (negative examples).

For the sample of lensed galaxies, we include the 11 bright {\it Planck\/} dusty GEMS, as well as H-ATLAS J114637.9-001132 (a strong gravitationally-lensed galaxy at $z\,{=}\,3.3$ discovered using the first {\it Planck\/} compact source catalogue; \citealt{FuLensed2012}) and HLS J091828.6+514223 (a bright gravitationally-lensed galaxy at $z\,{=}\,5.2$; \citealt{CombesLensed2012}). We also include a number of less bright gravitational lenses, which are crucial for identifying similar candidate galaxies in the SPIRE maps. Specifically, we include the two highest redshift lensed galaxies from \citet{NegrelloLenses2010}, namely SDP.81 at $z\,{=}\,3.04$ \citep{ALMASDP812015} and SDP.130 at $z\,{=}\,2.63$, as well as all of the lensed galaxies from \citet{WeissLensed2013} that have a peak flux density ${>}\,$100\,mJy at 350 or 500\,$\mu$m (16 galaxies total). We exclude the lower redshift lensed galaxies from \citet{NegrelloLenses2010} because they are below the redshift range ($z\,{\approx}\,2$--4) of galaxies expected in the PHz sample. We exclude the fainter lensed galaxies from \citet{WeissLensed2013} because they are virtually indistinguishable from non-lensed galaxies in the SPIRE maps (the threshold at 100\,mJy is motivated by e.g., \citealt{Paciga2009, NegrelloLenses2010, Nayyeri2016}). In other words, this is a cutoff in the magnification of the training lensed galaxies. The final training set thus contains a total of 31 confirmed high-$z$ strong gravitationally-lensed galaxies.

To create a sample of non-lensed star-forming galaxies, we turn to the multi-wavelength follow-up programmes of candidate protoclusters of star-forming galaxies from PlanckXXVII. \citet{MacKenzieSCUBA2017} and \citet{ChengSCUBA2020} followed up large samples of PHz sources with SCUBA-2 at 850\,$\mu$m and selected 45 sources as protocluster candidates, based on their overdensities at this wavelength. Although these sources are well studied, this does not guarantee that there are undiscovered lensed galaxies within these protocluster candidates. Again, motivated by the criteria used to select gravitationally-lensed galaxies, we do not include galaxies with flux densities ${>}\,100\,$mJy at 350/500\,$\mu$m in the unlensed training set (11 galaxies across the 45 sources). The flux densities of these individual galaxies are largely indistinguishable from the flux densities of individual galaxies in less extreme sources. The training set contains the remaining 388 non-obviously gravitationally-lensed galaxies from the protocluster candidate sources. For model validation, we randomly select 20\,per cent of the training set (84 galaxies) to be a test set, on which the model is tested but not trained.

\subsubsection{Neural-network model}
We construct our ML model using {\sc PyTorch} \citep{Paszke2019}. We find that a simple NN model consistently trains to about 100\,per cent accuracy on the test set, however, it does not consistently identify the same new candidate lensed galaxies each time it is trained. That is, the training data set is not sufficiently large or representative to fully constrain the weights and biases of the NN model, causing the random values in the initialized weights and biases to influence the NN's classification results. Even with small, single-layer models, the results of the NN are not fully consistent across multiple training attempts, despite consistently training to 100\,per cent accuracy on the test set. This is a common problem in ML that is typically overcome by using ensemble methods to combine the results of multiple ML models, making predictions more reliable \citep{OptizEnsemble1999, RokachEnsemble2010}. Fortunately, training a simple NN model is fast enough to make ensemble learning feasible (a simple single-layer NN take about 10\,s to train for 5000 epochs on a standard desktop). We adopt an ensemble learning technique in which we simultaneously train many NN models and use the average of their outputs for more consistent classification \citep{HansenNN1990}.

Our final ML classification model consists of 100 ensemble NNs. Each NN has three input nodes (corresponding to galaxy flux densities at 250, 350 and 500\,$\mu$m), one hidden layer of 20 nodes, and a single output node, for which an output of 1.0 corresponds to a lensed galaxy and 0.0 corresponds to an un-lensed star-forming galaxy. We adopt a standard rectified linear unit (ReLU; \citealt{NairReLU2010}) activation function for the hidden layer, a sigmoid (or logistic) activation function for the output layer, and the adaptive moment estimation optimizer (ADAM; \citealt{KingmaADAM2014}). We simultaneously train all 100 NNs over 5000 epochs (using the full training set every epoch) with a learning rate of 0.01, and average their outputs to obtain a final ML score. We use a standard, albeit arbitrary, cutoff at 0.5 to identify new lensed galaxies.

\subsubsection{New candidate lensed galaxies}
The ML model identifies a total of eight candidate gravitationally-lensed galaxies among the PHz HELP sources (see Table~\ref{HELPlensedtable}). As expected, the ML model confidently identifies G160 as a candidate gravitationally-lensed galaxy, which, as noted above, has been confirmed in the {\it Herschel\/} Stripe 82 field \citep{GeachG1602015, HarringtonG1602016}. G160 is easily identifiable without sophisticated methods due to its exceptional brightness, comparable to {\it Planck}'s Dusty GEMS (an order of magnitude above the other candidate galaxies in the PHz HELP sources), in addition to being the only notable galaxy in this PHz source. In contrast, most of the other identified candidate lensed galaxies are contained within PHz sources that consist of many SPIRE galaxies, including ${>}\,3\,\sigma$ overdensities G088 and G245. Notice that in the PHz HELP sources, we find only one exceptionally bright (peak flux density ${\gtrsim}\,300$\,mJy) strong gravitationally-lensed galaxy out of 187 sources (0.5\,per cent), whereas PlanckXXVII found a total of 12 exceptionally bright strong gravitational lenses (the 11 {\it Planck\/} GEMS and one previously discovered galaxy) out of 228 sources (5\,per cent). The larger number of bright lensed galaxies in PlanckXXVII can be explained by the inclusion of PCCS sources in the follow-up programmes and the bias towards targeting bright sources. As a result, the population of bright gravitationally-lensed galaxies (peak flux ${\gtrsim}\,300$\,mJy) in the PHz has been previously overestimated; we find them to be significantly more rare (perhaps 0.5\,per cent, as opposed to 5\,per cent). Of course there could still be a significant population of somewhat fainter gravitationally lensed galaxies in these catalogues.

Applying our ML model to galaxies within the PlanckXXVII sources, we find that we can identify 13 new candidate lensed galaxies from the PlanckXXVII sources (see Table~\ref{PlanckXXVIIlensedtable}). Additionally, we note that two PlanckXXVII maps contain very bright sources that lie outside of the IN region but still within 10\,arcmin of {\it Planck} position (we include them in Table~\ref{PlanckXXVIIlensedtable}). One of the best lensed candidates from the PlanckXXVII sources, G224, has recently been followed up with ALMA and confirmed as a new {\it Planck}-identified strong gravitationally-lensed galaxy (Fig.~\ref{G224ALMA}). The 870-$\mu$m ALMA image has spatially resolved G224, revealing a clear Einstein ring. This supports the suspected presence of gravitationally-lensed galaxies among the PHz sources that are less extreme than the {\it Planck\/} GEMS. Additionally, unlike the GEMS, G224 contains many SPIRE-detected galaxies, similar to many of the other candidates, rather than having a single dominant galaxy.

\begin{figure*}
\includegraphics[width=0.83\textwidth]{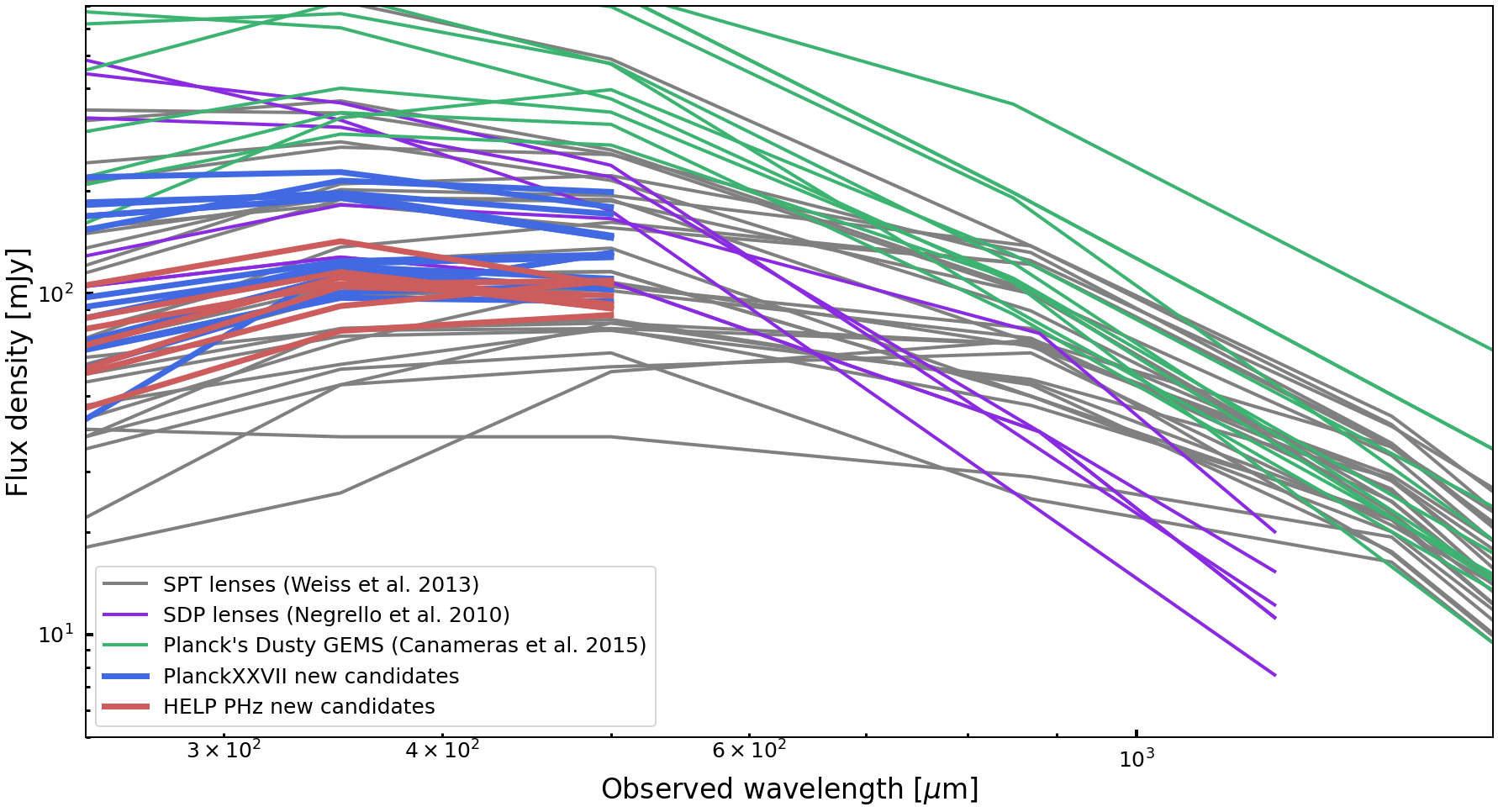}
\caption{Flux densities of the 7 new PHz HELP lensed candidates (red) and 13 new PlanckXXVII lensed candidates (blue) in comparison with other confirmed gravitationally-lensed galaxies from \citet{WeissLensed2013} (grey), \citet{NegrelloLenses2010} (purple) and {\it Planck}'s dusty GEMS (green). Although the new lensed candidates are less bright than {\it Planck}'s dusty GEMS, their flux densities are consistent with other high-$z$ gravitationally-lensed galaxies.}
\label{lensedSEDs}
\end{figure*}

The SEDs of the new candidates, both from the PHz HELP sources and the PlanckXXVII sources, are consistent with previously-discovered lensed galaxies (see Fig.~\ref{lensedSEDs}). In particular, the rising SEDs and peak flux densities of the new candidates are most comparable to the sample of lensed galaxies from \citet{WeissLensed2013}, which have an average redshift of 3.5. Their SEDs are also similar to the highest redshift lensed galaxies in \citet{NegrelloLenses2010}, namely SDP.81 and SDP.130, which were included in our training data set. The other lower-redshift lensed galaxies from \citet{NegrelloLenses2010} clearly peak at lower flux densities than these new candidates. Additional observations are required to determine whether the candidate lenses are true strong gravitationally-lensed galaxies, as well as to confirm their potentially high-redshift nature.

\subsubsection{500-$\mu$m threshold}
\label{simplethreshold}
As briefly mentioned above, previous studies searching for strong gravitationally-lensed galaxies in the submm took advantage of the steep number counts of non-lensed star-forming submm galaxies by adopting a strict flux-density selection threshold \citep{Coppin2006, Negrello2007}. In particular, a cutoff of 100\,mJy at 500\,$\mu$m has been used to search for candidate lensed galaxies in {\it Herschel} wide-field surveys \citep{NegrelloLenses2010, Nayyeri2016}. Due to the high efficiency of this threshold at selecting lensed galaxies, we implicitly relied on this threshold when constructing the training data set. Of course, this raises the question of whether the ML model picks up on additional features when identifying candidate lensed galaxies or simply reproduces the results of a 100-mJy threshold at 500\,$\mu$m.

Among the PHz HELP sources, a total of four galaxies (out of 1892) have a flux density above 100\,mJy at 500\,$\mu$m. The ML model, on the other hand, identifies eight candidate lensed galaxies, including all four galaxies with flux densities above 100\,mJy at 500\,$\mu$m. The additional four galaxies identified by the ML model are bright at 500\,$\mu$m and either have increasing SEDs or a flux density above 100\,mJy at 350\,$\mu$m (see Table~\ref{HELPlensedtable}). In the PlanckXXVII sources, a total of 30 galaxies (out of 1916) are brighter than 100\,mJy at 500\,$\mu$m, significantly more than the PHz HELP sources. Excluding previously discovered sources, 18 galaxies have flux density above 100\,mJy at 500\,$\mu$m and the ML model identifies a total of 13 new candidate lensed galaxies. There is significant overlap between the 18 galaxies with $S_{500}\,{>}\,100$\,mJy and the 13 candidate lensed galaxies from the ML model; 12/13 of the potential lensed galaxies have $S_{500}\,{>}\,100$\,mJy and the last galaxy has an increasing SED with $S_{500}$ just below $100$\,mJy. The six galaxies with $S_{500}\,{>}\,100$\,mJy that the ML model does not identify as candidate lensed galaxies have decreasing or flat SEDs (nonetheless, we include them in Table~\ref{PlanckXXVIIlensedtable}).

Because the ML model identifies candidate lensed galaxies with increasing SEDs and excludes some galaxies with decreasing/flat SEDs, the ML model is clearly more sophisticated than a simple threshold at 500\,$\mu$m; however, it is unclear whether this additional effort is worthwhile. When dealing with the bright galaxies within the PlanckXXVII sources, a simple threshold of 100\,mJy at 500\,$\mu$m identifies nearly all of the ML candidate galaxies, as well as a handful of additional galaxies with decreasing or flat SEDs. These galaxies could be gravitationally lensed of course (possibly at lower redshift), but one could add an additional colour cut (e.g., $S_{350}/S_{250}\,{>}\,1.2$) to the simple 500\,$\mu$m threshold to remove such galaxies if desired. For the generally less bright galaxies within the PHz HELP sources, we see more of an advantage to using the ML model, since it doubles the number of candidate lensed galaxies identified from four to eight. However, some of the additional galaxies would instead be picked up with the addition of other simple criteria (e.g., $S_{350}\,{>}\,100$\,mJy). As such, it is not clear if the increased sophistication and potential for identifying additional lensed galaxies with ML approaches is worth the required effort. Nonetheless, the classifications of our ML model are helpful for informing the choices of simple classification criteria (see below).

\subsubsection{New threshold selection criteria}
\label{manuallearning}

\begin{figure*}
\includegraphics[width=\textwidth]{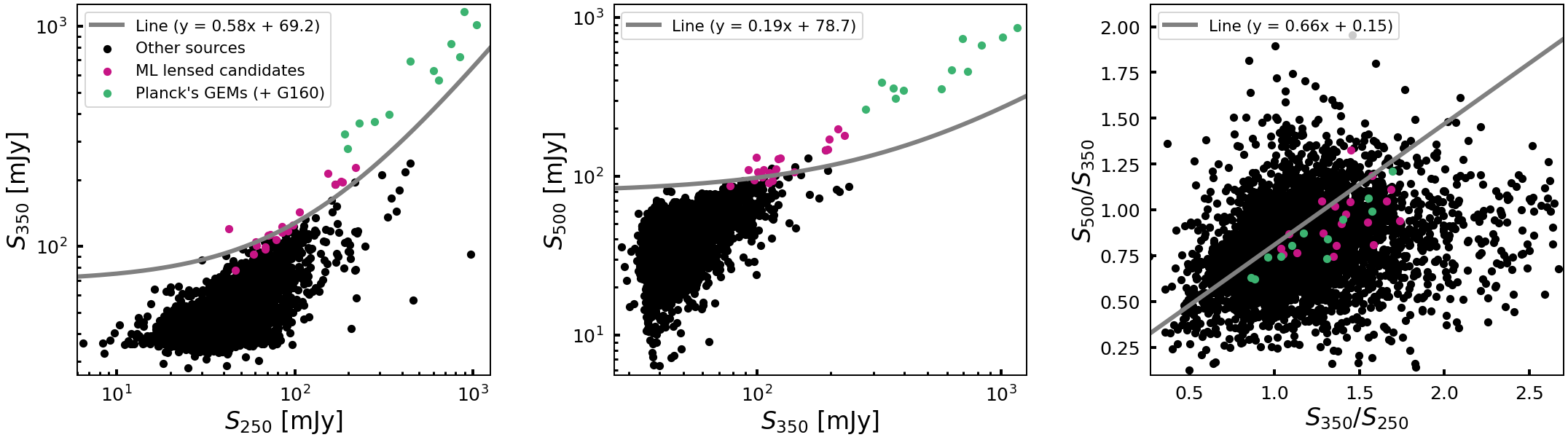}
\caption{Flux densities and flux density ratios for all the PHz HELP and PlanckXXVII sources, with {\it Planck}'s dusty GEMS highlighted in green and the new candidate lensed galaxies identified by the ML model highlighted in magenta. Optimized demarcation lines for reproducing the results of the ML model are shown in grey. The new candidate lensed galaxies cannot be identified based on SPIRE colours alone, but the ML model's classifications can be reproduced reasonably well with simple linear cuts in the $S_{350}$ versus $S_{250}$ and $S_{500}$ versus $S_{350}$ space plots.}
\label{manuallearningscatter}
\end{figure*}

One of the advantages of providing flux densities at 250, 350 and 500\,$\mu$m as inputs to the ML model, as opposed to the unprocessed images themselves, is the greater potential to interpret the classifications of the ML model and to tune more straight-forward selection criteria. Based on $S_{250}$, $S_{350}$ and $S_{500}$, {\it Planck}'s dusty GEMS are easily distinguished from non-lensed submm galaxies (see Fig.~\ref{manuallearningscatter}). Using flux ratios (i.e., colours) alone, there is no clear distinction, but considering both brightness and flux ratios, there is an obvious separation between {\it Planck}'s dusty GEMS and typical submm galaxies (see left and centre panels of Fig.~\ref{manuallearningscatter}). Furthermore, we see that the ML model identifies the galaxies that best follow the trends of {\it Planck}'s dusty GEMS in $S_{350}$ versus $S_{250}$ and $S_{500}$ versus $S_{350}$ at lower brightnesses.

In the hope of reproducing the classification done by the ML model with a threshold criteria, we introduce demarcation lines in $S_{350}$ versus $S_{250}$, $S_{500}$ versus $S_{350}$ and $S_{500}/S_{350}$ versus $S_{350}/S_{250}$ spaces. We optimize the slope and intercept of each demarcation line based on the classifications of the ML model. To incorporate both the accuracy and precision of the demarcation line criteria in searching for the best line, we optimize for the line's $F_1$ score (the harmonic mean of the precision and recall: $F_1 = 2/(\mathrm{recall}^{-1} + \mathrm{precision}^{-1})$). In other words, for each of the plots in Fig.~\ref{manuallearningscatter}, we search the two-dimensional parameter space to find the slope and intercept of the line that maximizes the $F_1$ score with respect to the ML model's classifications. As expected, the candidate lensed galaxies cannot be accurately distinguished from the non-lensed galaxies using their colours alone; although the candidate lensed galaxies are red in colour, so are the majority of non-lensed galaxies; the `optimal' line ($S_{500}/S_{350}\,{=}\,0.66\,S_{350}/S_{250}\,{+}\,0.15$) has $F_1$ score {<}\,0.5. For the absolute flux density plots, on the other hand, we see that the demarcation lines are able to separate the lensed galaxy candidates from the non-lensed galaxies much more effectively. The optimal demarcation line in $S_{350}$ versus $S_{250}$ space ($S_{350}\,{=}\,0.58\,S_{250}\,{+}\,69.2$\,mJy) classifies galaxies as lensed candidates/non-lensed galaxies with an $F_1$ score of 0.999, and similarly, the optimal line in $S_{500}$ versus $S_{350}$ space ($S_{500}\,{=}\,0.19\,S_{350}\,{+}\,78.7$\,mJy) also classifies galaxies with an $F_1$ score of 0.999. We define the threshold selection criteria as follows: galaxies that fall above the $S_{350}$ versus $S_{250}$ demarcation line or the $S_{500}$ versus $S_{350}$ demarcation line are identified as candidate lensed galaxies. We choose to identify candidate lensed galaxies that fall above either of the two demarcation lines because this slightly increases the $F_1$ score of the individual lines alone and results in a higher $F_1$ score than requiring galaxies to lie above both lines.

The threshold selection criteria lead to six candidate lensed galaxies among the PHz HELP sources, all of which were identified by the ML model. The criteria miss G245 due to its lower flux density at $350$ and $500$\,$\mu$m, which was likely identified by the ML model as a lensed galaxy based on its increasing SED, and narrowly miss G010 due to its relatively high flux density at $250$\,$\mu$m (note that these galaxies also did not have a high ML score). Among the PlanckXXVII sources, the threshold criteria identifies 15 candidate lensed galaxies, including all of the galaxies identified by the ML model, except G223 due to its lower flux density at $350$ and $500$\,$\mu$m. The other three galaxies identified by the threshold criteria all have $S_{500}\,{>}\,100$\,mJy; in fact, they are the three brightest galaxies at 500\,$\mu$m not identified as lensed galaxies by the ML model. Despite the simplicity of the criteria, two linear flux thresholds are able to reproduce the results of the ML model reasonably well, missing a few of the galaxies with increasing SEDs, while also identifying some additional plausible lensed galaxies. Furthermore, our criteria are less strict than a threshold of 100\,mJy at 500\,$\mu$m, potentially allowing for the detection of additional lensed galaxies that are fainter. Further follow-up observations are required to confirm the lensed nature of candidate galaxies identified by either the ML model or linear threshold criteria.

\section{Discussion}
\label{discussion}

\subsection{The nature of the PHz sources}
Now that we have investigated the resolved rest-frame far-IR properties of a representative subsample of PHz sources, we can discuss the overall statistical properties of the PHz catalogue, including some corrections to the conclusions drawn with the biased subsample analysed in PlanckXXVII.

In the preferentially-selected PlanckXXVII subsample, which had a bias towards high S/N sources with compact/regular shapes and included sources from the PCCS, 12 sources turned out to be ultra-bright strong gravitationally-lensed galaxies (5\,per cent of their subsample). Based on our more representative sample, we now see that the fraction of bright strong gravitational lenses (peak flux ${\gtrsim}\,300$\,mJy) in the PHz catalogue has been overestimated; we find only one such galaxy (a rate of 0.5\,per cent), namely G160. The larger fraction of bright strong gravitationally-lensed galaxies found in the PlanckXXVII sample suggests that such bright gravitational lenses are more easily found in the {\it Planck\/} compact source catalogues (see \citealt{Trombetti2021}), rather than the PHz catalogue.

On the other hand, we confirm that most PHz sources correspond to overdensities in SPIRE maps. This is particular illustrated through the stacking analysis, in which average stacks of the PHz HELP sources, similarly to the PlanckXXVII sources, show significant extended emission out to about 5\,arcmin, well beyond the SPIRE beam at all three observed wavelengths (with the most substantial differences at $500$\,$\mu$m). Additionally, we found a total of 40 sources having ${>}\,3\,\sigma$ overdensities of SPIRE galaxies, corresponding to 21\,per cent of the PHz HELP sources; this is comparable to the 25\,per cent of ${>}\,3\,\sigma$ overdensities found in the PlanckXXVII sources and includes a larger fraction of sources that are overdense at 350 and 500\,$\mu$m.

Despite the similarity in these results, we find significant differences between the PHz HELP and PlanckXXVII sources when we examine the SEDs of the galaxies found across the two subsamples. For our PHz HELP sources, we measure lower overdensities at 250 and 500\,$\mu$m when compared to the PlanckXXVII sources, with a slightly greater positive tail at 350\,$\mu$m (see Fig.~\ref{sourcedens}). This is related to the differences in colour seen between the galaxies in these two subsamples, for which we find larger values of $S_{350}/S_{250}$ in the PHz HELP galaxies and larger values of $S_{500}/S_{350}$ in the PlanckXXVII galaxies. This may indicate that galaxies found in the PlanckXXVII sources are at systematically higher redshift or contain systematically cooler dust temperatures, although we cannot provide any definitive conclusions because of the large uncertainties associated with the SPIRE photometric redshifts. Indeed, the distributions of photometric redshifts appear nearly identical for the two subsamples (see Fig.~\ref{zSFRdists}).

We {\it know\/} that the PHz sources are places on the sky corresponding to peaks in the CIB that are red. Hence we know that these are special directions, containing galaxies with red colours at these wavelengths. However, we would like to know whether the observed projected overdensities are predominantly protoclusters of galaxies at the same redshift or line-of-sight alignments of two or more groups of galaxies. Unfortunately, our current photometric redshifts are too uncertain to distinguish between these two situations. Instead, to make progress, we turn to a comparison with simulations. Specifically, we repeated the PHz selection algorithm on the Simulated Infrared Dusty Extragalactic Sky \citep[SIDES;][]{BetherminSIDES2017} simulation data. Briefly, SIDES simulated 2\,deg$^2$ of the extragalactic sky at the wavelengths observed by the SPIRE instrument (i.e., 250, 350 and 500\,$\mu$m), as well as 850\,$\mu$m, using a two-scenario model of star formation, where galaxies are either actively forming stars or quenched. Clustering was included by populating dark matter halos from dark matter-only simulations using an abundance matching technique (see \citealt{BetherminSIDES2017} for more details). We smoothed the SIDES catalogue to the 5\,arcmin resolution of {\it Planck}, and reproduced the 500-$\mu$m excess selection criteria used to generate the PHz catalogue (see \citealt{PlanckPHz2016}). Upon investigating the 30 brightest peaks in the resulting excess map, we did not find any strong evidence of large-scale clustering or line-of-sight alignments; instead, the selection criteria predominantly picked out individual galaxies bright at 500\,$\mu$m, with only mild clustering. This is likely caused by the small size of the SIDES simulation, preventing it from reproducing the rare objects that make up the PHz catalogue; the 2151 PHz sources were found across 10{,}700\,deg$^2$, meaning that their density is about 0.2\,deg$^{-2}$, and thus a simulation at least as large as $10\,{\rm deg}^2$ would be needed to find comparably rare objects.

Given the limitations of the SIDES simulation, we also perform tests using an analytical formalism as an alternative way to investigate potential line-of-sight alignments. Specifically, we employ the analytic model developed by \citet{Negrello2017}, which gives the luminosity function (LF) of clumps (either an isolated galaxy or a physically bound group of galaxies) at 545\,GHz based on galaxy evolution models and the infrared LFs of individual galaxies; this model is found to match observations well. Following the simple Monte Carlo simulations done by \citet{Negrello2017}, we draw clumps from the LF of their model, and distribute them randomly over 5{,}000\,deg$^2$ of a mock sky --- we expect this area to be large enough to generate the rare peaks in the CIB found by {\it Planck}, and we note that the 40 protocluster candidates found here were drawn from 1{,}270\,deg$^2$.

We smooth the mock sky with a Gaussian corresponding to the {\it Planck\/} beam of 5\,arcmin. From this smoothed map we select peaks with S/N$\,{>}\,$5 to mimic the PHz selection, where the noise is the rms of the map, that is, purely the confusion noise. Finally, from this sample, we select sources with 3\,$\sigma$ overdensities of individual galaxies brighter than ${\simeq}\,32$\,mJy at 545 GHz, corresponding to our selection of protocluster candidates as described in Sect.~\ref{PCcandidates}. The 500-$\mu$m excess selection used in \citet{PlanckPHz2016} to eliminate low-redshift sources and other potential contamination is not necessary here because the LF adopted for the test only includes clumps with redshifts between 2 and 4.

From this test we find about 220 simulated protocluster candidates (or line-of-sight alignments), selected in the same way as our protocluster candidate catalogue. This roughly matches the number of candidates projected for the given simulated area. Upon investigating these simulated protocluster candidates, we find that 96\,per cent of them contain two or more clumps aligned along the line-of-sight that are not physically connected to one another. Similarly, 81\,per cent are found to have more than three clumps aligned along the line-of-sight by chance. The average number of physically unbound clumps along the line-of-sight is about four, while that for random directions is about one.

This suggests that most of the PHz sources, and our protocluster candidates selected from that catalogue, may be line-of-sight combinations of star-forming galaxies at different redshifts, rather than single clumps of physically associated galaxies. This is similar to the conclusions drawn in \citet{Negrello2017} and \citet{Gouin2022}, which found that the {\it Planck}-derived SFRs for the PHz sources can be explained only by a dominant contribution from background/foreground interlopers, rather than by single haloes with very high SFRs. This conclusion, that there are foreground and background interlopers, is of course not very surprising. The more interesting question is whether there is a dominant star-forming protocluster along these special lines of sight. From the same simulations we find that the ratios of flux densities between the brightest and second brightest clumps along the selected lines of sight are much higher than that in random directions, with the brightest clump being a factor of approximately 2 (3) times brighter than the second brightest clump in 76 (60)\,per cent of the cases. The average flux density ratio between the brightest and second brightest clumps is about 4.4, in comparison with 1.5 for random directions. These results suggest that while the PHz sources and our protocluster candidates do tend to have line-of-sight interlopers, they still tend to be dominated by a significantly bright clump of galaxies at a single redshift, motivating continued follow-up studies with protocluster science in mind.

To summarize the conclusions of this section, we find that the PHz catalogue consists primarily of angular overdensities of $z\,{\sim}\,2$--3 galaxies, with only a very small fraction of bright strong gravitational lenses. Simulations suggest that there are multiple structures along each line of sight, but that they are often dominated by a single structure. In order to determine whether these spatial overdensities contain genuine protoclusters, further spectroscopic follow-up observations are needed. Our set of 40 ${>}\,3\,\sigma$ overdensities provides an optimal list for such studies, and should help us better understand how efficient CMB surveys are at finding the population of star-forming protoclusters around redshifts of 2.

\subsection{The spatial distribution of star-formation in $z\,{\sim}\,2$ protoclusters}

\begin{figure*}
\includegraphics[width=0.9\textwidth]{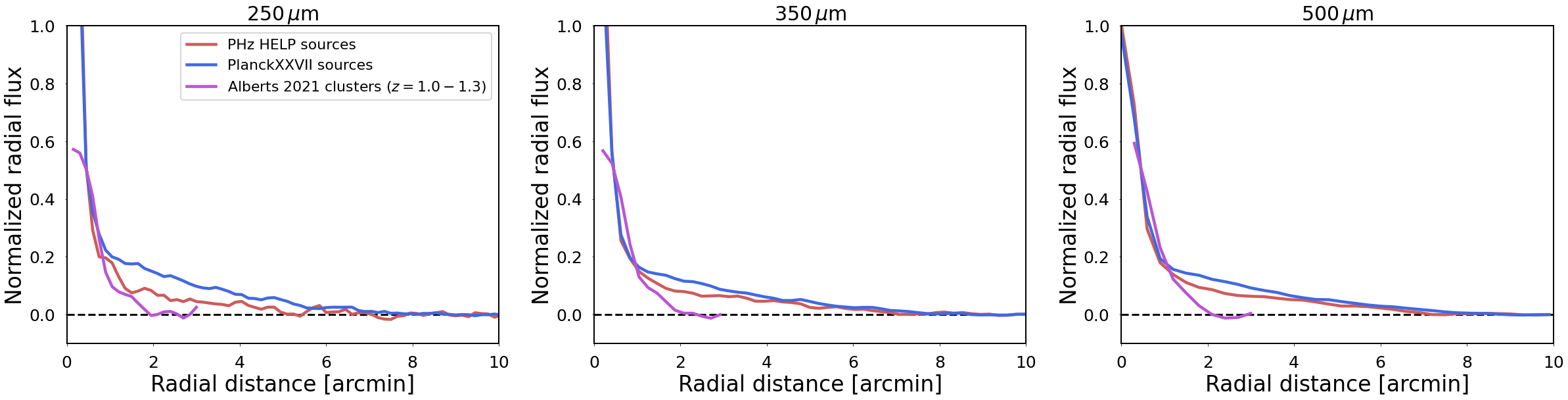}
\caption{Radial surface brightness profiles for the stacked PHz HELP sources, PlanckXXVII sources, and the $z\,{=}\,1.0$--1.3 clusters from \citet{Albertsclusters2020}. For consistency, we use the same bin widths as \citet{Albertsclusters2020} and normalize to 0.5 at 27.5\,arcsec. Both the PHz HELP sources and the PlanckXXVII sources are more extended than the lower-$z$ stacked clusters.}
\label{radialclustercomparison}
\end{figure*}

Current models of galaxy formation underestimate the SFRs observed in high-$z$ protoclusters \citep{LimProtoclusters2020}, including the PHz source G237 \citep{PollettaG2372021}. Even though simulations suggest that chance line-of-sight alignments are quite common among PHz sources, if each line-of-sight is at least dominated by one bright clump (as suggested above), then our {\it Herschel\/}-SPIRE observations may still be useful for constraining star-formation in typical protoclusters at $z\,{\sim}\,2$. In particular, in Section~\ref{radialfluxes} we stacked cutouts of the PHz HELP sources centered on the brightest galaxy detected at 350\,$\mu$m in each source. We expect that these brightest galaxies are part of the dominant clumps, and therefore our stacks may average out the fainter clumps along each line-of-sight.

Currently, statistical samples of protoclusters at higher redshift do not exist, but at $z\,{\approx}\,1$ \citet{Albertsclusters2020} studied 53 near-IR-selected galaxy clusters observed by {\it Herschel}-SPIRE at 250, 350 and 500\,$\mu$m. In Fig.~\ref{radialclustercomparison} we compare our stacked $z\,{\sim}\,2$ surface brightness profiles (as well as those obtained from PlanckXXVII for reference) to the $z\,{\approx}\,1$ stack from \citet{Albertsclusters2020} at 250, 350 and 500\,$\mu$m. Here we have followed \citet{Albertsclusters2020}, adopting the same bin sizes and arbitrarily normalizing the profiles to a value of 0.5 at 27.5\,arcsec.

Interestingly, we find that the average far-infrared extent of the PHz HELP sources is much larger than that of the average $z\,{\approx}\,1$ cluster; it is worth noting that angular diameter distances only differ by about 5\,per cent between redshift 1 and 2, so the observed differences in size are physical. We know that the average SFR density of the Universe peaked around $z\,{\approx}\,2$ \citep[e.g.,][]{behroozi2013,gruppioni2013,MadauDickinsonCosmicSF2014,ChiangProtoclusters2017,koprowski2017}, and therefore we might expect $z\,{\sim}\,2$ protoclusters to be more extended in star-formation as well. However, this is speculation, and the exact mechanisms driving any increase in SFR and size would need to be investigated in more detail.

Similarly, it is interesting to note that the PHz HELP sources are more centrally concentrated than the PlanckXXVII sources. Since our stacking procedure relied on centering each image on the brightest 350\,$\mu$m-detected galaxy, the centre of each stack is effectively a point source with a flux density given by the average flux density of the brightest PHz HELP/PlanckXXVII galaxies. We found that across the PHz HELP sources, the brightest central galaxy is relatively brighter than its surroundings when compared with the PlanckXXVII sources. This could come from the bias towards brighter sources in selecting the PlanckXXVII sample, which not only picked brighter central galaxies, but also sources with bright galaxies distributed throughout {\it Planck}'s 5\,arcmin beam. With a more representative sample of PHz targets, we find that typical sources are, in fact, more likely to contain bright central galaxies relative to the surrounding faint galaxies, as opposed to collections of relatively bright galaxies. This may indicate the presence of brightest cluster galaxies (BCGs) forming in protocluster cores, although further work is needed to test this speculation.

\section{Conclusions}
\label{conclusion}
We have used archival {\it Herschel}-SPIRE data to perform a follow-up study of 187 representative sources from the PHz catalogue that serendipitously fall in the 1270\,deg$^{2}$ of sky covered by the HELP fields. This expands on the {\it Herschel}-SPIRE follow up of 228 preferentially-selected PlanckXXVII sources, of which only 83 ended up in the final PHz catalogue. We are thus able to determine statistical properties of the PHz catalogue, whereas the follow-up observations of PlanckXXVII suffered from a bias towards bright, compact sources with regular shapes, making the sample statistically distinct from the PHz catalogue.

In our sample of 187 PHz HELP sources, we find only one (previously-identified) exceptionally bright (peak flux ${\gtrsim}\,300$\,mJy) strong gravitationally-lensed galaxy that is comparable to the 11 {\it Planck\/} GEMS found in PlanckXXVII. This implies that the proportion of exceptionally bright strong gravitationally-lensed galaxies in the PHz has been overestimated by a factor of about 10, and the true fraction of such galaxies in the PHz catalogue is likely to be {<}\,1\,per cent.

The other PHz HELP sources consist primarily of red galaxies, most of which are overdense when compared with random locations at 350 and 500\,$\mu$m, demonstrating that typical PHz sources are potential protoclusters. We investigated the average surface brightness profile of our representative sample of PHz sources using a stacking analysis, finding significant extended emission (out to about 5\,arcmin) relative to stacks around random galaxies on the sky. We also compared our surface brightness profiles to similar stacks of $z\,{\approx}\,1$ galaxy clusters selected in the near-IR, and found our sources to be much more extended.

In search of the most promising protocluster candidates, we identify a total of 40 PHz HELP sources that are ${>}\,3\,\sigma$ overdense at 250, 350 or 500\,$\mu$m (21\,per cent). For comparison with the PCCS, \citet{GreensladeHerschelPCCS2018} found 27 protocluster candidates that are ${>}\,3\,\sigma$ overdense out of a total of 354 PCCS sources (8\,per cent). The larger sample of significant overdensities in the PHz catalogue (most of which are at 350/500\,$\mu$m) compared with the PCCS supports the usefulness of the PHz catalogue in selecting protocluster targets. Due to the larger number of sources in the PCCS, the density on the sky of ${>}\,3\,\sigma$ overdense protocluster candidates in the PHz catalogue ($3.1\,{\times}\,10^{-2}$\,deg$^{-2}$) and the PCCS ($3.3\,{\times}\,10^{-2}$\,deg$^{-2}$) is comparable.

Although we find only one exceptionally bright (peak flux ${\gtrsim}\,300$\,mJy) lensed galaxy in the PHz HELP, we have explored the possibility of finding more typical strong gravitationally-lensed galaxies in the remaining sources. Our machine-learning approach identified a total of seven new candidate lensed galaxies among the PHz HELP sources whose SEDs are consistent with being high-$z$ gravitational lenses. We also looked for lensed galaxies in the PlanckXXVII sources that may have been fainter than the bright GEMS. The algorithm identified 13 additional candidate lensed galaxies, including the PHz source G224 that is now confirmed by ALMA to be a gravitationally-lensed galaxy.

Our results support the high-$z$ nature of the PHz catalogue and confirm that the vast majority of the PHz sources are spatial overdensities of galaxies at 350 and 500\,$\mu$m. According to our results, we expect that the full catalogue of 2151 objects in the PHz contain overdensities of star-forming galaxies and are therefore interesting for follow-up studies in their own right. Our subset of the best 40 protocluster candidates and 7$\,{+}\,$13 new strong gravitational lenses provide a tractable and representative sample ideal for follow-up surveys (e.g., with ALMA or \textit{JWST}) to further elucidate the nature of the PHz catalogue, and significantly bolster the number of known protoclusters and strong gravitational lenses.

\section*{Acknowledgements}
We acknowledge Marta Frias Castillo for work on an early version of the extended gravitational lens catalogue. We acknowledge the support of the Natural Sciences and Engineering Research Council of Canada (NSERC). SL acknowledges partial support from the Canadian Institute for Theoretical Astrophysics. This work is based on observations obtained with {\it Planck\/} (\url{http://www.esa.int/Planck}), an ESA science mission with instruments and contributions directly funded by ESA Member States, NASA and Canada. {\it Herschel\/} is an ESA space observatory with science instruments provided by European-led Principal Investigator consortia and with important participation from NASA. The Herschel Extragalactic Legacy Project, (HELP), is a European Commission Research Executive Agency funded project under the SP1-Cooperation, Collaborative project, Small or medium-scale focused research project, FP7-SPACE-2013-1 scheme, Grant Agreement Number 607254. This paper makes use of the following ALMA data: ADS/JAO.ALMA\#2019.1.01155.S. ALMA is a partnership of ESO (representing its member states), NSF (USA) and NINS (Japan), together with NRC (Canada), MOST and ASIAA (Taiwan), and KASI (Republic of Korea), in cooperation with the Republic of Chile. The Joint ALMA Observatory is operated by ESO, AUI/NRAO and NAOJ.

\section*{Data availability}
All of the data used in this paper is publicly available. The PlanckXXVII SPIRE maps are available through the ESA Herschel Science Archive (\url{http://archives.esac.esa.int/hsa/whsa}). The wide-field HELP maps can be downloaded from the Herschel Extragalactic Legacy Project Data site (\url{http://hedam.lam.fr/HELP}). Catalogues of the candidate protoclusters and graviationally-lensed galaxies are included in the Appendices below.

\bibliographystyle{mnras}
\bibliography{refs}

\appendix

\section{PHz HELP source catalogues}
\label{appendix1}
Here we provide source catalogues for the PHz protocluster candidates and candidate gravitationally-lensed galaxies in the HELP fields. The protocluster candidate catalogue (Table~\ref{protoclustertable}) consists of the 40 PHz HELP sources that are at least 3$\,\sigma$ overdense at 250, 350 or 500\,$\mu$m when compared with random locations in the Lockman-SWIRE field. The catalogue provides the source name in the PHz catalogue, the {\it Planck\/} position, the galaxy counts at 250, 350 and 500\,$\mu$m, and the overdensity contrasts at 250, 350 and 500\,$\mu$m (see Section~\ref{PCcandidates}). We include the galaxy counts because some protocluster candidates have fewer galaxies within a small contour area and others have many galaxies within a large contour area.

The source catalogue of candidate strong gravitational lenses among the HELP sources (Table~\ref{HELPlensedtable}) consists of the eight galaxies that were identified as potential lensed galaxies by the ML process (seven new sources and G160). The catalogue columns give the PHz source in which the galaxy was detected, the position of the candidate lensed galaxy in the source, the flux density at 250, 350 and 500\,$\mu$m, and the ML model score from 0.0 to 1.0.

\begin{table*}
\centering
\caption{Names, locations, galaxy counts and overdensity contrasts for candidate PHz HELP protocluster candidates that are ${>}\,3\,\sigma$ overdense at 250, 350 or 500\,$\mu$m. Note that a 3$\,\sigma$ overdensity corresponds to an overdensity contrast of 1.78, 1.96 and 4.21 at 250, 350 and 500\,$\mu$m, respectively.}
\begin{tabular}{rcccccccc}
 \hline
 Source name\phantom{\quad\ } & {\it Planck\/} RA & {\it Planck\/} Dec & $N_{250}$ & $N_{350}$ & $N_{500}$ &$\delta_{250}$ & $\delta_{350}$ & $\delta_{500}$\\
 & [J2000] & [J2000] & & &\\
 \hline
  PHz G245.14$-$82.47 & 01:21:09.79 & $-$30:56:46.68 & 27 & 32 & 16 & \pp0.82 & 1.88 & \pz4.79\\
  PHz G105.94$-$57.19 & 00:14:52.98 & $+$04:28:16.39 & 38 & 27 & 13 & \pp2.24 & 2.07 & \pz4.94\\
  PHz G288.30$+$88.95 & 12:50:15.36 & $+$26:06:43.56 & 16 & 20 & 13 & \pp1.36 & 2.93 & \pz9.27\\
  PHz G325.50$-$56.97 & 23:23:16.56 & $-$56:07:12.72 & 22 & 19 & \pz8 & \pp1.67 & 2.08 & \pz4.21\\
  PHz G332.15$-$80.25 & 00:28:06.98 & $-$35:31:10.92 & 19 & 18 & \pz5 & \pp1.64 & 2.34 & \pz2.73\\
  PHz G097.44$+$86.77 & 12:57:51.60 & $+$30:02:11.04 & 10 & 17 & \pz8 & \pp0.45 & 2.29 & \pz5.21\\
  PHz G112.56$-$61.51 & 00:31:44.08 & $+$00:57:40.70 & 12 & 17 & \pz8 & \pp0.67 & 2.15 & \pz4.95\\
  PHz G229.11$+$27.02 & 08:56:06.96 & $-$00:47:18.23 & 14 & 17 & \pz7 & \pp1.46 & 2.98 & \pz5.59\\
  PHz G113.76$-$64.06 & 00:35:26.74 & $-$01:28:16.64 & 19 & 16 & 11 & \pp0.95 & 1.20 & \pz5.06\\
  PHz G326.25$-$53.62 & 23:04:54.72 & $-$58:24:59.04 & 11 & 16 & 10 & \pp1.03 & 2.95 & \pz8.91\\
  PHz G088.87$-$61.67 & 23:49:41.76 & $-$02:57:00.86 & 14 & 16 & \pz6 & \pp3.66 & 6.11 & \pz9.71\\
  PHz G309.49$-$72.52 & 00:40:25.18 & $-$44:27:57.60 & 16 & 15 & 10 & \pp1.11 & 1.65 & \pz6.09\\
  PHz G131.19$-$60.58 & 01:07:37.92 & $+$02:01:46.20 & 13 & 14 & \pz8 & \pp0.80 & 1.59 & \pz4.95\\
  PHz G116.79$-$63.75 & 00:40:35.11 & $-$01:00:45.68 & \pz9 & 14 & \pz7 & \pp0.69 & 2.52 & \pz6.06\\
  PHz G315.61$-$55.34 & 23:52:59.76 & $-$60:19:39.00 & 15 & 14 & \pz9 & \pp1.08 & 1.59 & \pz5.70\\
  PHz G249.65$-$81.40 & 01:23:58.15 & $-$32:02:08.16 & 10 & 13 & \pz8 & \pp1.90 & 4.04 & 11.46\\
  PHz G325.40$-$84.13 & 00:40:48.29 & $-$32:31:50.17 & 12 & 13 & \pz8 & \pp1.29 & 2.31 & \pz7.17\\
  PHz G332.06$-$57.45 & 23:07:16.80 & $-$53:22:54.11 & 14 & 13 & \pz7 & \pp1.82 & 2.49 & \pz6.55\\
  PHz G103.70$-$60.58 & 00:14:10.87 & $+$00:53:43.20 & 12 & 13 & 11 & \pp0.29 & 0.87 & \pz5.35\\
  PHz G316.81$-$83.76 & 00:44:17.78 & $-$33:10:35.39 & \pz8 & 13 & \pz5 & \pp0.58 & 2.43 & \pz4.30\\
  PHz G012.00$-$67.27 & 23:09:34.56 & $-$32:55:09.48 & 14 & 13 & \pz8 & \pp1.24 & 1.77 & \pz5.85\\
  PHz G156.51$+$88.24 & 12:47:00.48 & $+$28:35:18.96 & \pz8 & 12 & \pz5 & \pp1.03 & 3.07 & \pz5.81\\
  PHz G092.24$-$65.65 & 00:02:35.86 & $-$05:43:47.75 & 12 & 12 & \pz8 & \pp0.67 & 1.22 & \pz4.95\\
  PHz G096.79$-$62.76 & 00:04:52.45 & $-$02:16:40.01 & \pz8 & 12 & \pz6 & \pp0.87 & 2.74 & \pz6.52\\
  PHz G259.75$-$84.40 & 01:09:19.27 & $-$31:08:12.84 & 16 & 12 & \pz7 & \pp1.58 & 1.58 & \pz5.05\\
  PHz G298.56$-$88.58 & 00:51:55.94 & $-$28:32:51.00 & 15 & 12 & \pz7 & \pp1.94 & 2.14 & \pz6.36\\
  PHz G012.90$-$66.24 & 23:04:31.92 & $-$32:41:45.95 & 10 & 11 & \pz4 & \pp1.05 & 2.00 & \pz3.39\\
  PHz G342.15$-$81.32 & 00:25:05.34 & $-$33:41:56.76 & 12 & 11 & \pz8 & \pp1.19 & 1.68 & \pz6.82\\
  PHz G117.68$-$57.72 & 00:40:11.30 & $+$05:02:13.60 & \pz9 & 11 & \pz2 & \pp1.41 & 2.93 & \pz1.87\\
  PHz G223.72$-$79.87 & 01:36:49.78 & $-$28:34:01.92 & \pz7 & 10 & \pz3 & \pp2.05 & 4.82 & \pz6.01\\
  PHz G006.02$-$65.14 & 23:01:30.96 & $-$35:40:04.44 & 10 & 10 & \pz4 & \pp1.48 & 2.31 & \pz4.31\\
  PHz G015.12$-$65.76 & 23:01:52.80 & $-$31:50:25.80 & \pz7 & 10 & \pz2 & \pp0.86 & 2.54 & \pz1.85\\
  PHz G355.11$-$84.41 & 00:30:56.42 & $-$30:27:54.00 & 14 & 10 & \pz5 & \pp1.87 & 1.74 & \pz4.50\\
  PHz G014.00$-$64.56 & 22:56:22.32 & $-$32:24:00.72 & \pz9 & 10 & \pz4 & \pp1.38 & 2.52 & \pz4.66\\
  PHz G007.50$-$64.16 & 22:56:12.24 & $-$35:14:09.60 & \pz8 & \pz9 & \pz6 & \pp1.19 & 2.29 & \pz7.82\\
  PHz G029.60$+$80.99 & 13:31:34.56 & $+$26:14:52.44 & \pz9 & \pz8 & \pz5 & \pp2.57 & 3.23 & \pz9.62\\
  PHz G317.59$-$54.94 & 23:43:22.56 & $-$60:12:41.04 & \pz8 & \pz8 & \pz9 & \pp0.11 & 0.48 & \pz5.70\\
  PHz G322.01$-$63.40 & 23:56:50.16 & $-$51:39:36.00 & 10 & \pz8 & \pz4 & \pp1.69 & 1.88 & \pz4.78\\
  PHz G242.28$-$81.81 & 01:24:43.99 & $-$30:53:49.20 & \pz7 & \pz8 & \pz4 & \pp1.88 & 3.40 & \pz7.83\\
  PHz G019.33$-$68.20 & 23:12:49.20 & $-$30:04:16.68 & \pz5 & \pz7 & \pz4 & \pp0.29 & 1.41 & \pz4.52\\
  $^{a}$PHz G358.75$-$82.52 & 00:22:31.59 & $-$31:08:39.84 & 13 & 19 & \pz7 & \pp0.40 & 1.74 & \pz3.05\\
  $^{a}$PHz G025.58$+$77.88 & 13:44:33.84 & $+$24:57:16.56 & 16 & 19 & \pz7 & \pp0.68 & 1.66 & \pz2.94\\
  $^{a}$PHz G026.02$-$85.28 & 00:30:17.78 & $-$27:35:48.84 & 10 & 15 & \pz7 & \pp0.14 & 1.28 & \pz3.28\\
  $^{a}$PHz G083.88$-$64.06 & 23:47:03.36 & $-$06:10:55.85 & 12 & 14 & \pz6 & \pp0.85 & 1.88 & \pz3.95\\
  $^{a}$PHz G010.97$-$78.47 & 00:01:33.29 & $-$30:52:58.44 & \pz6 & 14 & \pz6 & $-$0.17 & 1.59 & \pz3.46\\
  $^{a,b}$PHz G095.50$-$61.59 & 00:00:46.47 & $-$01:26:57.77 & 10 & 13 & \pz5 & \pp0.39 & 1.41 & \pz2.72\\
  $^{a}$PHz G344.06$+$89.34 & 12:53:22.32 & $+$26:37:55.56 & \pz8 & 13 & \pz3 & \pp0.28 & 1.78 & \pz1.58\\
  $^{a}$PHz G009.58$+$83.12 & 13:19:07.68 & $+$24:14:06.00 & \pz9 & 12 & \pz2 & \pp0.25 & 1.22 & \pz0.49\\
  $^{a}$PHz G110.89$-$63.19 & 00:29:50.64 & $-$00:49:30.02 & \pz6 & 11 & \pz2 & $-$0.07 & 1.29 & \pz0.67\\
  $^{a}$PHz G081.53$+$87.74 & 12:58:16.56 & $+$28:49:00.12 & 11 & 10 & \pz4 & \pp0.98 & 1.41 & \pz2.87\\
  $^{a}$PHz G303.03$+$86.33 & 12:51:27.84 & $+$23:27:44.64 & \pz8 & \pz8 & \pz2 & \pp0.83 & 1.44 & \pz1.45\\
 \hline
\end{tabular}
\begin{tablenotes}
  \small
  \item {${}^{a}$ Source identified due to $3\,\sigma$ overdensity of red galaxies (see Section~\ref{3sigmacuts}).}
  \item {${}^{b}$ Studied with follow-up observations spanning from the optical to the sub-mm in \citet{Flores-CachoG952016}.}
\end{tablenotes}
\label{protoclustertable}
\end{table*}

\begin{table*}
\centering
\caption{Names, locations and flux densities for the candidate lensed galaxies in the PHz catalogue that happen to fall in the HELP fields, with the mean score of the 100 ensemble NNs.}
\begin{tabular}{rcccccc} 
 \hline
 Source name\phantom{\quad\ } & RA & Dec & $S_{250}$ & $S_{350}$ & $S_{500}$ & ML score\\
 & [J2000] & [J2000] & [mJy] & [mJy] & [mJy]\\
 \hline
  $^{a}$PHz G160.57$-$56.79 & 02:09:41.19 & $+$00:15:57.70 & 891 $\pm$ 10 & 1165 $\pm$ 11 & 857 $\pm$ 10 & 1.00\\
  PHz G088.87$-$61.67 & 23:49:39.97 & $-$02:55:54.29 & \pz58 $\pm$ 8\pz & \pz\pz92 $\pm$ 8\pz & 109 $\pm$ 8\pz & 0.98\\
  PHz G115.61$-$59.35 & 00:36:27.67 & $+$03:20:51.64 & 106 $\pm$ 8\pz & \pz142 $\pm$ 8\pz & 106 $\pm$ 8\pz & 0.92\\
  PHz G222.75$-$55.98 & 03:25:01.90 & $-$27:33:08.76 & \pz79 $\pm$ 6\pz & \pz107 $\pm$ 6\pz & 109 $\pm$ 6\pz & 0.91\\
  PHz G325.82$-$62.47 & 23:44:07.68 & $-$51:32:24.65 & \pz60 $\pm$ 6\pz & \pz105 $\pm$ 6\pz & \pz99 $\pm$ 7\pz & 0.84\\
  PHz G358.75$-$82.52 & 00:22:32.56 & $-$31:11:36.99 & \pz71 $\pm$ 4\pz & \pz112 $\pm$ 6\pz & \pz91 $\pm$ 7\pz & 0.63\\
  PHz G245.14$-$82.47 & 01:20:52.35 & $-$30:55:41.35 & \pz46 $\pm$ 7\pz & \pz\pz78 $\pm$ 6\pz & \pz87 $\pm$ 7\pz & 0.59\\
  PHz G010.71$-$70.28 & 23:24:01.94 & $-$32:55:30.32 & \pz85 $\pm$ 5\pz & \pz116 $\pm$ 5\pz & \pz93 $\pm$ 6\pz & 0.50\\
 \hline
\end{tabular}
\begin{tablenotes}
  \small
  \item {${}^{a}$ Previously discovered bright strong gravitationally-lensed galaxy \citep{GeachG1602015, HarringtonG1602016}.}
\end{tablenotes}
\label{HELPlensedtable}
\end{table*}

\section{Candidate lensed galaxies in the PlanckXXVII sources}
\label{appendix2}
PlanckXXVII identified 11 new exceptionally bright (peak flux ${\gtrsim}\,300$\,mJy) candidate gravitationally-lensed galaxies among the PlanckXXVII sources, which have all since been confirmed, but there may be less obvious undiscovered lensed galaxies in the PlanckXXVII sources. As with the PHz HELP sources, we use our ML model to identify more dim candidate lensed galaxies. The source catalogue of new candidate PlanckXXVII lensed galaxies (Table~\ref{PlanckXXVIIlensedtable}) provides the {\it Herschel}-SPIRE source name, the position of the candidate lensed galaxy, the flux densities at 250, 350 and 500\,$\mu$m, and the ML score. One of the best candidates, G224, has recently been followed up with ALMA and confirmed as a strong gravitational lens system (Fig.~\ref{G224ALMA}).

\begin{table*}
\centering
\caption{Names, locations and flux densities for the candidate lensed galaxies in the PlanckXXVII sources that may have been missed by PlanckXXVII, with the mean score of the 100 ensemble NNs.}
\begin{tabular}{rcccccc} 
 \hline
 Source name\phantom{\quad\ } & RA & Dec & $S_{250}$ & $S_{350}$ & $S_{500}$ & ML score\\
 & [J2000] & [J2000] & [mJy] & [mJy] & [mJy]\\
 \hline
  $^{a}$PLCK\_G224.6$-$40.7 & 04:33:19.24 & $-$25:46:32.22 & 154 $\pm$ 6 & 215 $\pm$ 5 & 199 $\pm$ 6 & 1.00\\
  PLCK\_G003.7$-$63.3 & 22:53:38.49 & $-$37:03:22.31 & \pz85 $\pm$ 6 & 123 $\pm$ 5 & 128 $\pm$ 6 & 1.00\\
  PLCK\_G026.6$+$74.4 & 13:59:38.09 & $+$24:23:16.15 & 169 $\pm$ 5 & 191 $\pm$ 4 & 146 $\pm$ 5 & 1.00\\
  PLCK\_G058.5$+$64.6 & 14:44:26.76 & $+$35:12:35.47 & 220 $\pm$ 6 & 227 $\pm$ 5 & 180 $\pm$ 5 & 0.99\\
  PLCK\_G016.2$+$54.0 & 15:12:17.55 & $+$12:36:27.95 & \pz61 $\pm$ 6 & 101 $\pm$ 5 & 106 $\pm$ 5 & 0.99\\
  PLCK\_G120.4$+$51.6 & 13:06:14.00 & $+$65:25:17.97 & 182 $\pm$ 6 & 197 $\pm$ 5 & 171 $\pm$ 6 & 0.99\\
  PLCK\_G186.3$-$72.7 & 01:56:33.97 & $-$18:28:36.80 & \pz73 $\pm$ 6 & 113 $\pm$ 5 & 105 $\pm$ 6 & 0.98\\
  PLCK\_G319.7$-$45.3 & 22:43:26.17 & $-$67:23:58.05 & \pz97 $\pm$ 5 & 124 $\pm$ 5 & 130 $\pm$ 6 & 0.96\\
  PLCK\_G143.6$+$69.4 & 12:10:04.73 & $+$46:06:26.77 & \pz69 $\pm$ 6 & \pz99 $\pm$ 5 & 132 $\pm$ 5 & 0.96\\
  PLCK\_G176.6$+$59.0 & 10:37:07.85 & $+$41:25:32.20 & 185 $\pm$ 6 & 195 $\pm$ 5 & 148 $\pm$ 6 & 0.92\\
  PLCK\_G256.8$-$69.3 & 02:09:29.96 & $-$40:07:26.51 & \pz43 $\pm$ 5 & 120 $\pm$ 5 & 111 $\pm$ 5 & 0.63\\
  PLCK\_G223.3$+$61.4 & 10:52:10.68 & $+$18:52:54.40 & \pz68 $\pm$ 5 & \pz97 $\pm$ 5 & \pz95 $\pm$ 6 & 0.62\\
  PLCK\_G239.0$+$69.8 & 11:35:55.44 & $+$17:04:50.51 & \pz91 $\pm$ 6 & 118 $\pm$ 5 & 103 $\pm$ 5 & 0.60\\
  $^{b}$PLCK\_G242.3$-$40.7 & 04:43:45.39 & $-$38:59:00.45 & 375 $\pm$ 6 & 368 $\pm$ 6 & 264 $\pm$ 6 & 1.00\\
  $^{b}$PLCK\_G180.1$+$65.1 & 11:04:27.10 & $+$38:12:32.99 & 267 $\pm$ 8 & 269 $\pm$ 7 & 284 $\pm$ 7 & 0.92\\
  $^{c,d}$PLCK\_G191.3$+$62.0 & 10:44:38.47 & $+$33:51:03.63 & 101 $\pm$ 6 & 115 $\pm$ 5 & 112 $\pm$ 6 & 0.20\\
  $^{c,d}$PLCK\_G106.8$-$83.3 & 00:43:25.58 & $-$20:42:00.88 & 161 $\pm$ 3 & 162 $\pm$ 3 & 129 $\pm$ 3 & 0.11\\
  $^{c}$PLCK\_G107.6$+$36.9 & 16:07:22.53 & $+$73:47:02.67 & 217 $\pm$ 6 & 195 $\pm$ 5 & 109 $\pm$ 5 & 0.02\\
  $^{c}$PLCK\_G160.7$+$41.0 & 09:07:27.92 & $+$56:09:00.57 & 111 $\pm$ 5 & 119 $\pm$ 5 & 101 $\pm$ 5 & 0.01\\
  $^{c,d}$PLCK\_G112.4$+$45.8 & 14:17:07.61 & $+$69:32:27.35 & 162 $\pm$ 6 & 143 $\pm$ 5 & 114 $\pm$ 6 & 0.00\\
  $^{c}$PLCK\_G280.2$-$77.8 & 01:15:01.27 & $-$38:10:32.77 & 172 $\pm$ 6 & 152 $\pm$ 5 & 101 $\pm$ 6 & 0.00\\
 \hline
\end{tabular}
\begin{tablenotes}
  \small
  \item {${}^{a}$ New confirmed strong gravitationally-lensed galaxy (see Fig.~\ref{G224ALMA}).}
  \item {${}^{b}$ Bright galaxy located outside the IN region but still within 10\,arcmin of the {\it Planck} position.}
  \item {${}^{c}$ Galaxy identified with criteria from \citet{NegrelloLenses2010} ($S_{500}\,{>}\,100$\,mJy).}
  \item {${}^{d}$ Galaxy also identified with simple selection criteria (see Section~\ref{manuallearning}).}
\end{tablenotes}
\label{PlanckXXVIIlensedtable}
\end{table*}

\end{document}